\documentclass[journal]{IEEEtran}
\usepackage{cite}
\usepackage{amsmath,amssymb,amsfonts}
\usepackage{siunitx}
\usepackage{algorithmic}
\usepackage{graphicx}
\usepackage{optidef}
\usepackage{enumitem} % 用于自定义 enumerate 环境
\usepackage{float}
\usepackage{svg}  % For handling SVG files
\usepackage{float} % 在导言区添加这个包
\usepackage{placeins} % 用于插入 \FloatBarrier
\usepackage[ruled,vlined]{algorithm2e}
\usepackage{amsmath}

\usepackage{multirow}
\usepackage{textcomp}
\usepackage{xcolor}
\usepackage{subcaption}
\usepackage{bm}
\def\BibTeX{{\rm B\kern-.05em{\sc i\kern-.025em b}\kern-.08em
    T\kern-.1667em\lower.7ex\hbox{E}\kern-.125emX}}

\ifCLASSINFOpdf
\else
\fi
\begin{document}
\title{AGENS: Agentic Link Construction for Environment and Intent Aware 6G Communication}
\author{Zhaoyang Li, Shangzhuo Xie, Qianqian Yang
        % <-this % stops a space
\IEEEcompsocitemizethanks{
\IEEEcompsocthanksitem Part of this work has been accepted for presentation at the 2026 IEEE International Symposium on Personal, Indoor and Mobile Radio Communications (PIMRC), Singapore.
\IEEEcompsocthanksitem Zhaoyang Li, Shangzhuo Xie and Qianqian Yang are with the College of Information Science and Electronic Engineering, Zhejiang University, Hangzhou,
China. (e-mails: \{zhaoyangli, sz\_xie, qianqianyang20\}@zju.edu.cn).

\IEEEcompsocthanksitem This work is partly supported by the National Key R\&D Program of China under Grant No. 2024YFE0200802, by the National Natural Science Foundation of China (NSFC) under Grant Nos. 62293481, and 62571487, by the Zhejiang Provincial Natural Science Foundation of China under Grant No. LZ25F010001. (Corresponding author: Qianqian Yang.)

}
}

\markboth{Journal of \LaTeX\ Class Files,~Vol.~14, No.~8, August~2015}%
{Shell \MakeLowercase{\textit{et al.}}: Bare Demo of IEEEtran.cls for IEEE Journals}



\maketitle

\begin{abstract}
The emergence of diverse intelligent services is driving sixth-generation (6G) networks toward an intelligent communication ecosystem with increasingly complex operating scenarios. However, existing AI-based communication methods mainly focus on isolated tasks or individual physical-layer modules, which is insufficient for AI-native 6G systems that must jointly perceive dynamic environments, understand specific user intents, coordinate multiple coupled PHY-layer modules, and remain compatible with existing communication infrastructures. To address this gap, this paper, for the first time, explores \emph{automatic environment- and intent-aware full-link construction}, which aims to generate executable PHY-layer configurations according to channel conditions and service requirements.
We propose AGENS, an agentic communication decision-making framework for AI-native 6G link construction. Rather than replacing the existing PHY-layer protocol stack with a black-box end-to-end model, AGENS operates on top of the modular communication framework and jointly optimizes end-to-end PHY-layer configurations. Specifically, AGENS encodes CSI and natural-language user instructions, aligns communication states with user intents through cross-modal fusion, and leverages a pretrained LLM for intent-aware reasoning. An LSTM-based sequential action generator further captures dependencies among coupled sub-actions and produces structured full-link strategies. To enable stable training in the high-dimensional discrete action space, AGENS adopts a two-stage learning framework, where heuristic supervised initialization is followed by multi-objective reinforcement learning to jointly optimize reliability, throughput, and power consumption. Simulation results under diverse channel conditions and user intents show that AGENS rapidly generates near-optimal full-link configurations without the time-consuming online search required by heuristic algorithms. The dataset and code are available at: https://github.com/LZhaoyang/AGENS.
\end{abstract}
% Extensive experimental results demonstrate that CSI-ALM achieves a 1 dB gain over state-of-the-art deep learning methods. 

% \fi
\IEEEpeerreviewmaketitle

\section{Introduction}
% and "HIS" in caps to complete the first word
\IEEEPARstart{R}{E}{C}{E}{N}{T} advances in artificial intelligence (AI) have significantly accelerated the evolution of wireless communications. Learning-based methods have achieved strong performance in a variety of communication tasks, including channel estimation, channel prediction, beam prediction, beamforming, and resource allocation, etc\cite{b1,b2}. However, as communication scenarios become increasingly heterogeneous and service requirements more diverse, optimizing a single module in isolation is often no longer sufficient. For instance, a high-throughput link may require the joint adaptation of modulation order, coding rate, power compensation, channel estimation, and precoding, whereas a reliability-oriented link may favor more conservative modulation and coding together with robust channel estimation and equalization. This challenge is expected to become even more prominent in sixth-generation (6G) networks, where communication systems must simultaneously support personalized user requirements, adapt to dynamic environments, and coordinate multiple coupled physical-layer modules toward global link-level objectives\cite{b3,b4}. These trends call for a unified AI-driven framework that goes beyond isolated task-specific optimization while remaining compatible with existing 5G and beyond physical-layer infrastructures.

Current 5G and beyond communication systems still rely largely on a modular physical-layer design. A complete communication link typically consists of channel coding and decoding, coding-rate selection, modulation and demodulation, power control, channel estimation, equalization, and MIMO precoding\cite{b5}. These functions are supported by mature communication techniques, such as LDPC\cite{b6} and polar codes\cite{b7} for channel coding, adaptive modulation and coding for link adaptation, LS\cite{b8} or MMSE\cite{b9} for channel estimation, linear equalizers for signal recovery, and ZF\cite{b10} or RZF\cite{b11} for precoding. In practical, the configurations of these modules are usually selected through predefined rules, lookup tables, or threshold-based mechanisms designed for specific channel conditions and service objectives\cite{b12,b13}. Although AI has also been introduced to improve some of these modules, most existing approaches still optimize individual functions separately. As a result, they have limited flexibility in capturing cross-module dependencies and adapting the overall link configuration to changing environments and diverse user intents.

%Recently, AI techniques have also been introduced to optimize these modules, usually by training task-specific predictors, classifiers, or policy networks for a particular communication function. However, most existing solutions still follow a single-module, single-objective, or rule-driven optimization paradigm, which makes the overall link configuration relatively rigid when facing diverse user intents and changing communication environments.

This limitation suggests that AI-native communication should not be viewed merely as applying AI to a collection of isolated tasks. Instead, a truly AI-native 6G system should jointly incorporate environment perception, channel-state awareness, user-intent understanding, cross-module PHY coordination, and compatibility with existing modular communication stacks. Large language models (LLMs) offer a promising foundation in delivering this goal. Owing to their strong capabilities in semantic understanding, instruction following, and contextual reasoning, LLMs can interpret high-level user requirements expressed in natural language and translate them into communication-oriented decision preferences, such as throughput enhancement, reliability improvement, or energy saving\cite{b14,b15,b16,b17,b18}. Moreover, when combined with multimodal learning techniques, LLMs can jointly process textual instructions and channel state information (CSI), allowing the decision process to account for both user requirements and wireless environments\cite{b19,b20}. Rather than replacing mature PHY modules, an LLM can therefore act as an agentic coordinator that reasons over performance trade-offs and produces structured decisions compatible with practical PHY implementations.

%Integrating LLMs with established communication technologies provides a practical path toward this vision. Rather than replacing mature PHY-layer modules, the LLM acts as an agentic coordinator that interprets user intents, incorporates CSI, reasons over performance trade-offs, and generates structured decisions compatible with practical PHY implementations. This LLM-empowered design enables adaptive, intent-aware, and executable full-link PHY construction beyond static rule matching and task-specific AI optimization.

However, existing LLM-based and wireless foundation model approaches still fall short of this goal. Recent studies such as WirelessGPT and WiFo demonstrate that pretrained models can support multiple downstream tasks or improve wireless representation learning\cite{b40a,b41,b42}. Nevertheless, these methods mainly serve as task-specific predictors, single-module optimizers, or general representation learners. Even when they consider multiple tasks, they do not explicitly address end-to-end link construction across multiple coupled PHY modules conditioned jointly on CSI and user requirements. This gap motivates the need for a new framework that can directly generate executable full-link configurations with both environment awareness and intent awareness.

Motivated by the above observations, we propose AGENS, an agentic communication decision-making framework for environment- and intent-aware full-link construction.  To the best of our knowledge, \textit{AGENS is the first framework that explicitly studies end-to-end PHY-link construction that is jointly aware of the wireless environment through CSI and of user intent through natural-language requirements.}
Specifically, AGENS generates executable PHY configurations that adapt to diverse channel conditions while satisfying personalized user requirements. 
To support closed-loop strategy evaluation and policy optimization, we first construct a digital-twin-inspired communication link simulator that mirrors the modular PHY-layer transmission process and provides link-level feedback. Based on this simulator, AGENS learns to optimize full-link configurations through interaction with the communication environment. 
To jointly model user intent and channel state, we design a cross-modal fusion module that aligns linguistic instructions with CSI. 
To address the large joint action space, AGENS adopts a structured sequential action-generation architecture that decomposes the full-link strategy into coupled PHY sub-actions, together with an LSTM-based state modeling component to capture dependencies among them. Finally, we develop a two-stage training pipeline: 
(i) heuristic exploration is used to construct expert-like samples, followed by supervised learning for strong policy initialization; (ii) multi-objective reinforcement learning is employed to fine-tune the policy by jointly optimizing BER, throughput, and power consumption. 
The main contributions of this paper are summarized as follows:
\begin{itemize}
    \item We propose AGENS, an agentic communication decision-making framework for environment- and intent-aware full-link construction. To the best of our knowledge, this is the first study to explicitly investigate agentic end-to-end PHY-link construction. By building a digital-twin-inspired communication link simulator, AGENS enables closed-loop strategy evaluation and policy optimization over modular PHY-layer configurations.
    
    \item We design a cross-modal fusion module to jointly model CSI and user intents expressed in natural language, enabling aligned reasoning over wireless conditions and service requirements. We further introduce an LSTM-based state modeling module to capture causal dependencies among coupled PHY sub-actions, thereby mitigating the modeling difficulty induced by the high-dimensional discrete action space.
     
    \item We develop a warm-start two-stage reinforcement learning framework tailored to AGENS. In Stage I, heuristic exploration is used to expand the experience buffer, followed by supervised learning to obtain a strong policy initialization. In Stage II, the policy is fine-tuned using PPO with an approximate entropy reward, improving decision quality while maintaining training stability.

    \item Simulation results demonstrate that AGENS can adaptively construct robust and efficient link strategies for diverse user requirements and challenging channel conditions, while rapidly generating near-optimal communication-link configurations without the time-consuming search required by heuristic algorithms.

\end{itemize}

The remainder of this paper is organized as follows. Section II reviews related work. Section III introduces the considered communication link model and formulates the decision-making problem. Section IV presents the proposed AGENS in detail. Section V describes the proposed reinforcement learning algorithm with an approximate entropy reward. Section VI reports and discusses the simulation results. Finally, Section VII concludes the paper.

\textbf{Notation:} Bold lowercase letters denote vectors, and bold uppercase letters denote matrices, while scalars are represented by plain letters. For example, for a vector $\bm{u}$, the $j$-th component is written as $u_j$. For a matrix $\bm{U}$, $\bm{U} \in \mathbb{R}^{\alpha \times \beta}$ specifies that $\bm{U}$ has dimensions $\alpha \times \beta$.

\section{Related Work}

\subsection{Traditional Communication Technologies}

Conventional wireless communication systems are built on a modular physical-layer design, where each functional block is independently designed and optimized according to well-established communication principles. For channel coding, LDPC and polar codes have been widely adopted in modern communication standards due to their strong error-correction capability and practical decoding efficiency\cite{b6,b7}. For link adaptation, adaptive modulation and coding (AMC) adjusts the modulation order and coding rate according to CSI or SNR, thereby balancing spectral efficiency and transmission reliability under time-varying channels\cite{b21}. For MIMO transmission, linear precoding and beamforming methods, such as ZF, RZF, and WMMSE, are commonly used to suppress inter-user interference and improve spectral efficiency\cite{b10,b11,b22}. In addition, channel estimation and equalization techniques, such as LS, MMSE, ZF, and LMMSE, are used to recover channel information and mitigate channel distortion at the receiver\cite{b8,b9,b23,b24}. 

Although these techniques form the foundation of current 5G and beyond systems, their configuration is still largely module-centric. In practical system, coding, modulation, power control, channel estimation, equalization, and precoding are typically selected using hand-crafted rules, lookup tables, or threshold-based switching mechanisms. Such designs are efficient in well-calibrated scenarios, but they rely heavily on scenario-specific tuning and often require re-calibration when channel conditions or service objectives change. More importantly, the coupled effects among PHY modules are difficult to capture under isolated design, which limits the ability of conventional systems to achieve globally coordinated full-link optimization.

\subsection{AI-Empowered Wireless Communications}

With the rapid development of deep learning, AI has been extensively applied to wireless communication. Early studies showed that neural networks can learn communication functions directly from data and approximate or complement conventional physical-layer processing under certain channel models~\cite{b25,b26}. In OFDM systems, deep learning has been used for channel estimation and signal detection, where neural networks learn to recover transmitted symbols from received pilots and data without relying entirely on explicitly derived estimators\cite{b27}. For interference management and resource control, learning-to-optimize methods have been developed to approximate iterative optimization algorithms and thereby accelerate power allocation and resource management decisions\cite{b28}. In massive MIMO and millimeter-wave systems, deep neural networks have been applied to channel estimation, beam prediction, and beamforming optimization by learning spatial and temporal channel features from large-scale datasets\cite{b29}. Sequence models such as RNNs, LSTMs, GRUs, and Transformers have been adopted for channel prediction, where historical CSI is used to infer future channel states and mitigate channel aging\cite{b30,b31,b32}. Reinforcement learning has also been introduced for user association, power control, and resource allocation in heterogeneous cellular networks and mobile edge computing systems\cite{b33}. 

These studies demonstrate that AI can significantly improve individual communication functions. However, most of them still focus on a single task, a single module, or a single optimization objective. As a result, they do not directly address the cross-module full-link decision problem required by AI-native 6G systems, where multiple coupled PHY modules must be coordinated under dynamic environments and diverse user requirements.

\subsection{LLMs Empowered Wireless Communications}

LLMs have recently attracted increasing interest in wireless communications due to their strong representation learning, contextual reasoning, and generalization capabilities\cite{b34}. One line of work applies pretrained LLMs to specific communication tasks. For channel prediction, LLM4CP incorporates task-specific embedding layers for frequency-domain and angular-domain CSI while keeping the LLM backbone largely frozen~\cite{b35}. CSI-LLM further aligns channel prediction with the autoregressive next-token generation mechanism of LLMs, enabling variable-length CSI inputs and multi-step prediction~\cite{b36}. For beam prediction, LLM-based methods convert beam management information into textual or prompt-like representations and use the LLM to improve robustness and generalization compared with conventional sequence models~\cite{b37}. For beamforming optimization, BERT4Beam serializes CSI into token sequences and reformulates beamforming as a sequence modeling problem, while task embeddings are introduced to indicate different optimization objectives so that a single model can handle multiple beamforming tasks~\cite{b38}. LLMs have also been used for resource allocation, where fine-tuned language models generate allocation policies with performance comparable to reinforcement learning-based methods under limited training data~\cite{b39}.

Another line of research develops wireless foundation models for general-purpose communication representation learning and multi-task adaption. Large Wireless Model (LWM) learns universal channel embeddings by reconstructing masked channel patches, aiming to provide task-independent channel representations~\cite{b40}. WirelessGPT jointly considers the time, space, and frequency dimensions of wireless channels and constructs a generative pretrained multi-task framework for communication, perception and task adaptation~\cite{b41}. In ~\cite{b42}, a multi-task wireless foundation model combines contrastive learning and masked autoencoding to learn channel representations transferable across multiple downstream tasks. These works demonstrate the potential of pretrained and foundation models in wireless communications. 

Nevertheless, existing LLM-based and foundation-model-based methods still mainly operate as task-specific predictors, single-module optimizers, or general representation learners. Even when multiple tasks are considered, they do not explicitly address end-to-end link construction across multiple coupled PHY modules conditioned jointly on CSI and user requirements. This limitation motivates the framework proposed in this paper, which targets environment- and intent-aware full-link construction over existing modular PHY components. 

\section{System model}
\subsection{Modular PHY-Layer Communication System}

We consider a 3GPP NR-compliant modular PHY-layer communication system, where the end-to-end transmission link is constructed by a set of functional communication modules~\cite{b43}.
Specifically, the considered link includes channel coding and decoding, modulation and demodulation, MIMO precoding, power control, channel estimation, and equalization.
Each module selects one strategy from its predefined candidate set, and the joint selection across all modules forms an executable end to end transmission configuration. The overall PHY-layer configuration is represented as a structured action vector:
\begin{equation}
\bm{a} = [a_{\mathrm{mod}}, a_{\mathrm{rate}}, a_{\mathrm{pc}}, a_{\mathrm{enc}}, a_{\mathrm{ce}}, a_{\mathrm{eq}}, a_{\mathrm{prec}}],
\end{equation}
where \(a_{\mathrm{mod}}\), \(a_{\mathrm{rate}}\), \(a_{\mathrm{pc}}\), \(a_{\mathrm{enc}}\), \(a_{\mathrm{ce}}\), \(a_{\mathrm{eq}}\), and \(a_{\mathrm{prec}}\) denote the selected strategies for modulation, coding-rate selection, power compensation, channel coding, channel estimation, equalization, and precoding, respectively.

Given the transmitted bit sequences \(\bm{x}\), the end-to-end transmission process under strategy \(\bm{a}\) and CSI \(\bm{H}\) can be modeled as:
\begin{equation}
    \hat{\bm{x}} = f(\bm{x}; \bm{a}, \bm{H}),
\end{equation}
where \(\hat{\bm{x}}\) denotes the recovered data at the receiver.

The performance of the modular PHY-layer communication system is evaluated using three metrics: bit error rate (BER), effective transmission rate, and power compensation overhead.
Let \(\bm{x}=[x_1,x_2,\ldots,x_{N_{\mathrm{bit}}}]\) and 
\(\hat{\bm{x}}=[\hat{x}_1,\hat{x}_2,\ldots,\hat{x}_{N_{\mathrm{bit}}}]\) 
denote the transmitted and recovered bit sequences, respectively.
The BER is calculated as:
\begin{equation}
    \mathrm{BER}
    =
    \frac{1}{N_{\mathrm{bit}}}
    \sum_{n=1}^{N_{\mathrm{bit}}}
    \mathbf{1}\left(x_n \neq \hat{x}_n\right),
\end{equation}
where \(\mathbf{1}(\cdot)\) denotes the indicator function, and 
\(N_{\mathrm{bit}}\) is the total number of bits used for BER calculation.

The effective transmission rate is defined as the ratio between the number of information bits and the number of transmitted modulation bits:
\begin{equation}
    R = \frac{K_{\mathrm{info}}}{K_{\mathrm{mod}}},
\end{equation}
where \(K_{\mathrm{info}}\) denotes the number of original information bits and \(K_{\mathrm{mod}}\) denotes the number of bits after modulation.

For power control, the selected strategy introduces an additional transmit-power offset \(\Delta P_{\mathrm{dB}}\) in dB. The corresponding power overhead is measured in the logarithmic domain as
\begin{equation}
    P_{\mathrm{extra}} = \ln\left(10^{\Delta P_{\mathrm{dB}}/10}+\epsilon\right),
\end{equation}
where \(\epsilon\) is a small positive constant introduced for numerical stability.

For the considered modular PHY-layer communication system, we implement the end-to-end transmission process using Sionna-RT, which provides a practical link-level simulation platform for strategy evaluation and policy optimization.

\subsection{AGENS Input--Output Formulation}

The proposed AGENS is a structured policy generation model that maps channel observations and user instructions to a complete PHY-layer configuration. Specifically, the input to AGENS is defined as:
\begin{equation}
\bm{s} = \{\bm{H}, \bm{T}\},
\end{equation}
where \(\bm{H}\) denotes the CSI and \(\bm{T}\) denotes the textual user instruction describing the communication requirement. Here, \(\bm{H} \in \mathcal{O}_{\mathrm{csi}}\) and \(\bm{T} \in \mathcal{O}_{\mathrm{text}}\), where \(\mathcal{O}_{\mathrm{csi}}\) and \(\mathcal{O}_{\mathrm{text}}\) denote the channel-observation space and textual-instruction space, respectively.

The output of AGENS is a structured action \(\bm{a}\in\mathcal{A}\), where \(\mathcal{A}\) denotes the predefined full-link action space formed by the candidate strategy sets of all configurable PHY modules. Accordingly, the overall decision process of AGENS can be formulated as:
\begin{equation}
\pi_{\theta} : \mathcal{O}_{\mathrm{csi}} \times \mathcal{O}_{\mathrm{text}} \rightarrow \mathcal{A},
\end{equation}
where \(\pi_{\theta}\) denotes the policy parameterized by \(\theta\). Therefore, AGENS learns a multimodal mapping from channel observations and user-intent descriptions to a structured full-link action:
\begin{equation}
\bm{a} = \pi_{\theta}(\bm{H}, \bm{T}).
\end{equation}

\subsection{Reinforcement Learning-Based Decision Modeling}

The joint strategy selection across communication modules is modeled as a reinforcement learning (RL)-based decision process. We define the decision process as
\begin{equation}
M = \{\mathcal{S}, \mathcal{A}, r\},
\end{equation}
where \(\mathcal{S}\) denotes the state space, \(\mathcal{A}\) denotes the action space, and \(r: \mathcal{S} \times \mathcal{A} \rightarrow \mathbb{R}\) represents the reward function.

The objective is to learn an optimal policy \(\pi_{\theta}: \mathcal{S} \rightarrow \mathcal{A}\) that maximizes the expected reward:
\begin{equation}
    \max_{\pi_{\theta} \in \Pi} \;
    \mathbb{E}_{\bm{s} \sim p_{\mathcal{S}},\, \bm{a} \sim \pi_{\theta}(\cdot|\bm{s})}
    \big[r(\bm{s}, \bm{a})\big],
\end{equation}
where \(p_{\mathcal{S}}\) denotes the state distribution over \(\mathcal{S}\), \(\bm{s} \in \mathcal{S}\) denotes the state, and \(\bm{a} \in \mathcal{A}\) denotes the action. Here, \(\pi_{\theta}(\bm{a}|\bm{s}) \in [0,1]\) represents the probability that policy \(\pi_{\theta}\) selects action \(\bm{a}\) under state \(\bm{s}\).
% In our framework, the proposed  model receives both textual information and CSI as inputs, and outputs the optimal strategy for each module in the physical-layer communication chain.  Accordingly, the policy parameterized by $\theta$ can be expressed as:
% \begin{equation}
%     \pi_{\theta}: \mathcal{O}_{\text{text}} \times \mathcal{O}_{\text{CSI}} \rightarrow \bm{a},
% \end{equation}
% where $\mathcal{O}_{\text{text}}$ denotes the textual observation space, $\mathcal{O}_{\text{CSI}}$ represents the channel state information space, and $\bm{a}$ corresponds to the set of strategies for all physical-layer modules. 

\section{Proposed Method}
In this section, we present the architecture of AGENS, as illustrated in Fig. \ref{fig1}. AGENS consists of four components: a CSI encoder and text encoder for multimodal perception, a cross-modal fusion module for aligning channel states with user intents, a pretrained language-model backbone for intent-aware representation learning, and an LSTM-based sequential action generator for structured PHY-layer strategy generation. The details of each component are introduced as follows.

\subsection{Multimodal Encoders}

\begin{figure*}[htbp]
    \centering
    \includegraphics[width=0.98\textwidth]{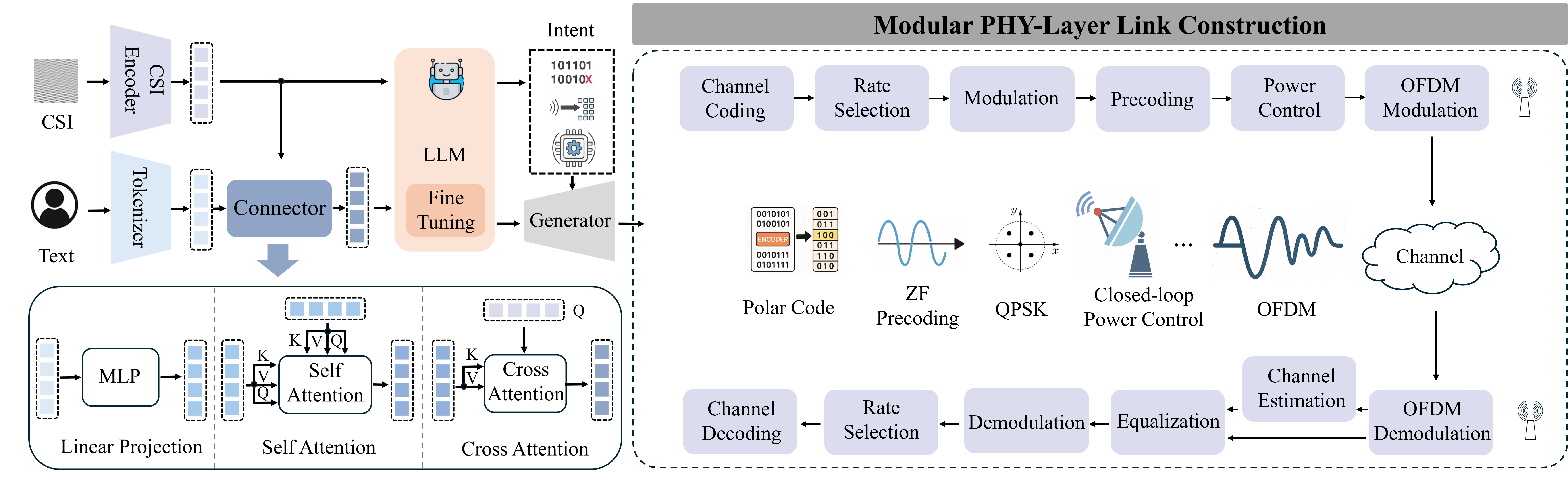} 
    \caption{Architecture of the proposed AGENS framework.}
    \label{fig1}
\end{figure*}

As illustrated in Fig.~\ref{fig1}, the multimodal encoder of AGENS consists of a CSI encoder and a text encoder, which extract channel-aware representations and semantic-aware textual representations, respectively.

The input CSI matrix is denoted by
\begin{equation}
\bm{H} \in \mathbb{C}^{L \times F},
\end{equation}
where \(L\) and \(F\) denote the temporal length and feature dimension of CSI observations, respectively. To facilitate transformer-style processing, the CSI sequence is first serialized through a temporal patching operation:
\begin{equation}
\bm{Y} = \text{Patch}(\bm{H}),
\end{equation}
where \(\bm{Y} \in \mathbb{R}^{2F \times N \times T'}\) and \(T'=\left\lceil \frac{L}{N}\right\rceil\).

Following our previous work~\cite{b36}, a Cross Spatial Self-Attention (CSSA) module is employed to model spatial-frequency correlations and inter-channel interference patterns. The refined CSI representation is expressed as
\begin{equation}
\overline{\bm{Y}}
=
\text{Reshape}
\left(
\text{CSSA}
\left(
\bm{Y}
\right)
\right),
\end{equation}
where \(\overline{\bm{Y}}\in\mathbb{R}^{2F\times L}\) denotes the enhanced CSI feature representation after spatial attention and structural reshaping.

To align the CSI representation with the semantic embedding space of the downstream LLM backbone, a linear projection layer is further applied:
\begin{equation}
\bm{Y}_{\mathrm{CSI}}
=
\text{Linear}
\left(
\overline{\bm{Y}}
\right),
\end{equation}
where \(\bm{Y}_{\mathrm{CSI}}\in\mathbb{R}^{L\times d}\) represents the CSI embedding in the latent semantic space with embedding dimension \(d\).

Meanwhile, the textual instruction \(\bm{T}\) is tokenized and embedded through the pretrained tokenizer and embedding layer of the LLM:
\begin{equation}
\bm{Y}_{\mathrm{text}}
=
\text{Embed}
\left(
\text{Tokenizer}(\bm{T})
\right),
\end{equation}
where \(\bm{Y}_{\mathrm{text}}\in\mathbb{R}^{L_t\times d}\) denotes the semantic embedding of the user instruction and \(L_t\) represents the text sequence length.

\subsection{Cross-Modal Fusion}
Directly feeding long textual instructions into the LLM incurs substantial computational overhead and may introduce redundant semantic information. To address this issue, we introduce a lightweight connector module before the LLM to compress and filter textual representations conditioned on CSI. First, a linear projection layer is applied to reduce the sequence length of the textual embeddings:
\begin{equation}
    \bm{Z}_{\text{text}} = \text{Linear}(\bm{Y_{\text{text}}}) ,
\end{equation}
where $\bm{Z}_{\text{text}} \in \mathbb{R}^{L_2 \times d}$ denotes the compressed textual representation.

Next, a self-attention mechanism is applied to capture semantic dependencies among tokens and aggregate global textual information:
\begin{equation}
    \tilde{\bm{Z}}_{\text{text}} =
    \text{Self-Attention}
    ({\bm{Z}}_{\text{text}}).
\end{equation}
Finally, inspired by the Q-Former structure, we introduce a cross-modal
attention mechanism to align CSI features with textual semantics\cite{li2023blip2}. Specifically, the CSI representation serves as the query, while the textual embeddings act as keys and values. The cross-modal attention operation is given by

\begin{equation}
 \tilde{\bm{{\mathit{Y}}}}_{\text{text}}= \text{Softmax}\left(\frac{Q K^T}{\sqrt{c}}\right)V,    
\end{equation}where
\begin{equation}
Q = \bm{{\mathit{Y}}}_{\text{CSI}}W_Q, \quad K = \tilde{\bm{Z}}_{\text{text}}W_K, \quad V = \tilde{\bm{Z}}_{\text{text}}W_V,    
\end{equation}
where $\mathbf{W}_Q$, $\mathbf{W}_K$, and $\mathbf{W}_V \in \mathbb{R}^{d \times d}$ denote learnable projection matrices. The scalar $c$ is a scaling factor. 

\subsection{LLM Backbone for Intent-Aware Reasoning}

After obtaining the CSI-aligned textual representation, we concatenate them with the CSI embeddings to construct a unified multimodal input sequence:
\begin{equation}
    \bm{X}^{0} = \text{Concat}[\tilde{\bm{{\mathit{Y}}}}_{\text{text}}; \bm{{\mathit{Y}}}_{\text{CSI}}],
\end{equation}
which \(\bm{X}^{0}\in\mathbb{R}^{2L\times d}\)  denotes the fused multimodal input to the LLM backbone.

The backbone network adopts a pretrained Transformer-based architecture composed of stacked self-attention layers and feed-forward networks. Given the input representation $\bm{X}^{l}$ at the $l$-th Transformer layer, the hidden representation is updated as:
\begin{equation}
    \bm{X}^{l+1} 
    =
    \text{FFN}
    \left(
    \text{MHA}
    \left(
    \bm{X}^{l}
    \right)
    \right),
\end{equation}
where $\mathrm{MHA}(\cdot)$ and $\mathrm{FFN}(\cdot)$ denote the multi-head self-attention module and the position-wise feed-forward network, respectively.

By progressively refining the multimodal representation across layers, the LLM captures high-order dependencies between physical-layer environments and semantic communication requirements. The final hidden representation generated by the LLM backbone is given by:
\begin{equation}
    \bm{{\mathit{Y}}}_{\text{LLM}}
    =
    \text{LLM}
    \left(
    \bm{X}^{0}
    \right),
\end{equation}
where $\bm{{\mathit{Y}}}_{\text{LLM}} \in \mathbb{R}^{2L \times d}$ denotes the refined multimodal feature representation.

\subsection{LSTM-Based Sequential Action Generator}
\begin{figure}[htbp]
    \centering
    \includegraphics[width=0.48 \textwidth]{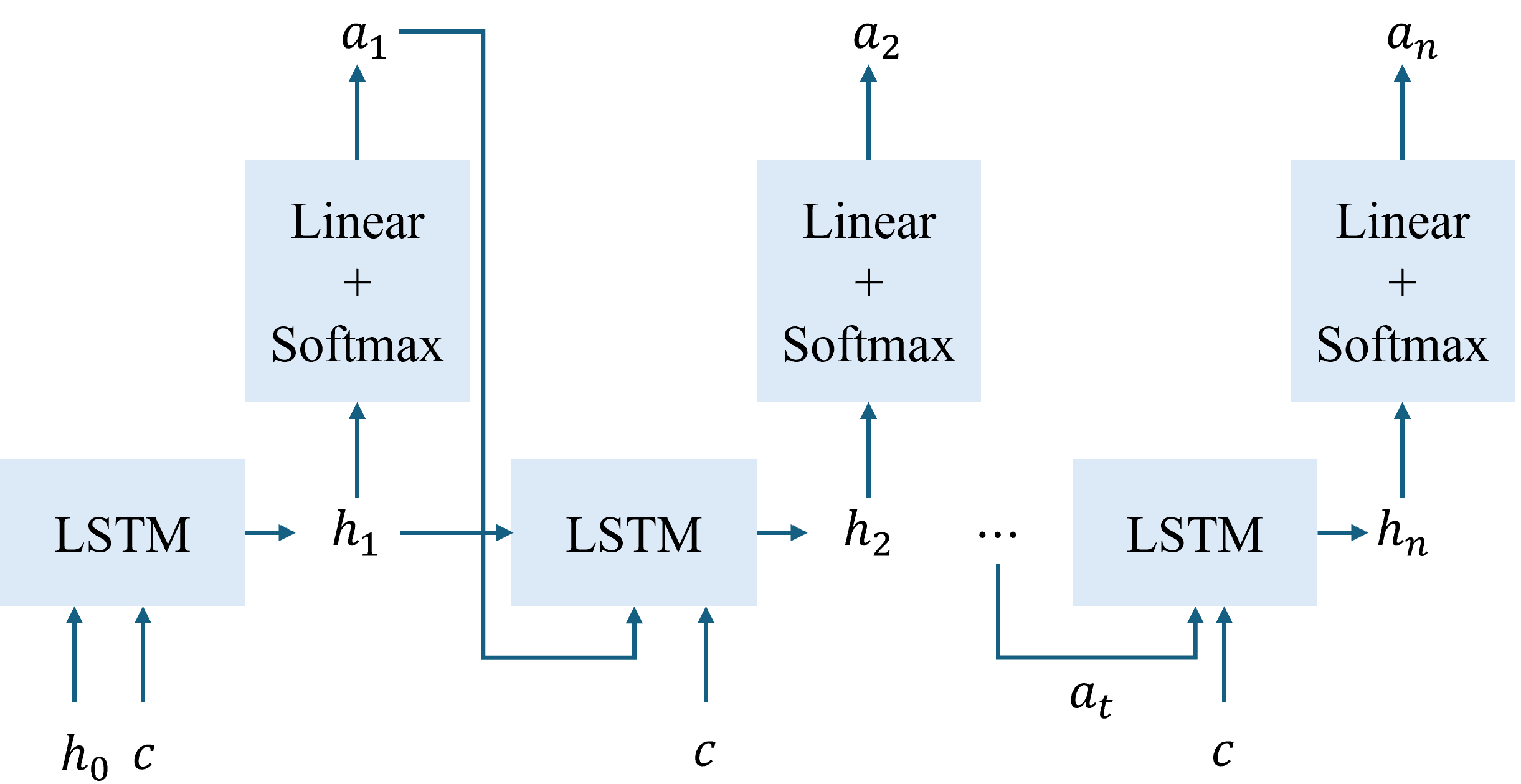} 
    \caption{LSTM-Based Sequential Action Generator.}
    \label{fig2}
\end{figure}
To model the dependencies among PHY modules and handle the structured action space efficiently, we adopt an LSTM-based sequential action generator rather than using independent actor heads for different sub-actions, as shown in Fig.~\ref{fig2}.

Let the overall physical-layer configuration be defined on a multidimensional discrete action space:
\begin{equation}
\mathcal{A}=\mathcal{A}_1 \times \mathcal{A}_2 \times \cdots \times \mathcal{A}_{n},
\end{equation}
where each sub-action \(a_i \in \mathcal{A}_i\) corresponds to the strategy selected for the \(i\)-th PHY module. Instead of modeling the joint policy over \(\mathcal{A}\) directly, we factorize it into a sequence of conditional distributions:
\begin{equation}
g_{\psi}(\bm{a} \mid \bm{{\mathit{Y}}}_{\text{LLM}})
=
\prod_{i=1}^{n}
g_{\psi}
\left(
a_i \mid \bm{{\mathit{Y}}}_{\text{LLM}}, a_1, \ldots, a_{i-1}
\right),
\end{equation}

Specifically, the multimodal representation generated by the LLM is first aggregated into a compact state embedding:
\begin{equation}
\bm{z}
=
\text{Pool}
\left(
\bm{{\mathit{Y}}}_{\text{LLM}}
\right),
\end{equation}
which is used to initialize the sequential action generation process.

The initial hidden state of the LSTM is given by:
\begin{equation}
\bm{h}_0 = \phi(\bm{z}),
\end{equation}
where $\phi(\cdot)$ denotes a learnable linear projection.

At generation step $i$, the LSTM takes the previous hidden state $\bm{h}_{i-1}$ and the embedding of the previously generated action $a_{i-1}$ as input:
\begin{equation}
\bm{h}_i
=
\text{LSTM}
\left(
\bm{h}_{i-1},
\text{Embed}(a_{i-1})
\right),
\end{equation}

The hidden state $\bm{h}_i$ is then mapped to a categorical distribution over the $i$-th action dimension:
\begin{equation}
g_{\psi}
\left(
a_i \mid \bm{{\mathit{Y}}}_{\text{LLM}}, a_1, \ldots, a_{i-1}
\right)
=
\text{Softmax}
\left(
W_i \bm{h}_i + b_i
\right),
\end{equation}

The resulted sub-action $a_i$ which maximizes $g_{\psi}(a_i \mid \bm{{\mathit{Y}}}_{\text{LLM}}, a_1, \ldots, a_{i-1})$ is then fed into the next decoding step. After all $n$ sub-actions are generated, the complete PHY link is obtained as:
\begin{equation}
\bm{a} = \{a_1, a_2, \ldots, a_{n}\},
\end{equation}

This autoregressive generation mechanism avoids enumerating the full joint action space while preserving dependencies among PHY-layer sub-actions.

\section{Training Method}
\begin{figure}[htbp]
    \centering
    \includegraphics[width=0.48\textwidth]{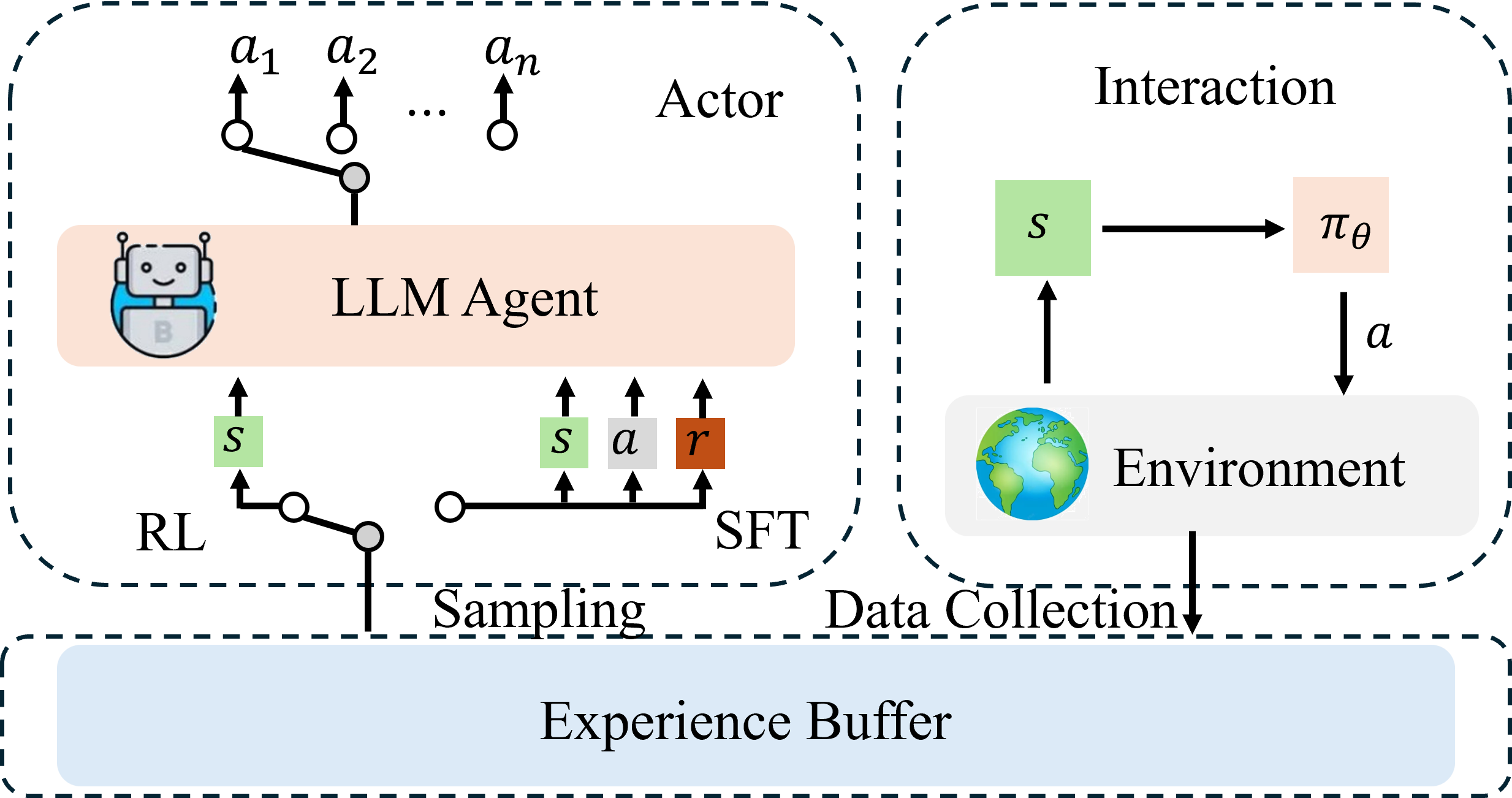} 
    \caption{Two-stage training method.}
    \label{fig3}
\end{figure}
In this section, we formulate intent-aware full-link configuration as a single-step reinforcement learning problem and develop the two-stage policy-optimization framework shown in Fig.~\ref{fig3}. The first stage initializes AGENS by supervised fine-tuning on heuristic expert samples, and the second stage improves the initialized policy using PPO with an entropy regularization term.

\subsection{Reinforcement Learning Formulation}

The link configuration problem is modeled as a one-shot RL decision process, where the agent jointly configures all PHY-layer modules according to the communication environment and the user requirement.

For the $t$-th sampled decision instance, the environment state is defined as
\begin{equation}
\bm{s}_t = \{\bm{H}_t, \bm{T}_t\},
\end{equation}
where $\bm{H}_t$ denotes the CSI and $\bm{T}_t$ represents the user instruction describing the communication preference. The index $t$ denotes a sampled configuration instance rather than a multi-step temporal transition.

Following the structured action formulation described in Section~III,
AGENS generates a complete PHY-layer configuration
\begin{equation}
\bm{a}_t=[a_{t,1},a_{t,2},\ldots,a_{t,n}],
\end{equation}
where $a_{t,i}\in\mathcal{A}_i$ denotes the sub-action for the $i$-th PHY module and $\mathcal{A}=\mathcal{A}_1\times\cdots\times\mathcal{A}_n$. To efficiently model this high-dimensional structured action space, the policy is parameterized as the sequentially factorized distribution
\begin{equation}
\pi_{\theta}(\bm{a}_t|\bm{s}_t)
=
\prod_{i=1}^{n}
\pi_{\theta}
\left(
a_{t,i}\mid
\bm{s}_t, a_{t,1:i-1}
\right),
\end{equation}
where $a_{t,1:i-1}$ denotes the previously generated sub-action sequence
before the $i$-th decoding step. Here, $\pi_{\theta}$ represents the
overall AGENS policy network, which consists of the multimodal encoders,
the cross-modal fusion module, the LLM backbone, and the LSTM-based
sequential action generator introduced in Section~IV. The LLM backbone is
fine-tuned using LoRA to adapt the pretrained language model to the
communication-link construction task. 

After executing action $\bm{a}_t$, the communication simulator evaluates
the resulting link performance and returns three metrics defined in
Section~III, namely $\mathrm{BER}_t$, $R_t$, and
$P_{\mathrm{extra},t}$. The scalar reward is defined as the weighted utility
of these normalized metrics:
\begin{equation}
\begin{aligned}
r_t
&=
r(\bm{s}_t,\bm{a}_t) \\
&=\; w_{\mathrm{BER}}\left(1-\eta_{\mathrm{BER}}(\mathrm{BER}_t)\right)
+ w_R\eta_R(R_t) \\
&+ w_P\left(1-\eta_P(P_{\mathrm{extra},t})\right),
\end{aligned}
\end{equation}
where $w_{\mathrm{BER}}$, $w_R$, and $w_P$ denote the weights for
reliability, throughput, and energy efficiency, respectively, with
$w_{\mathrm{BER}}+w_R+w_P=1$. The functions
$\eta_{\mathrm{BER}}(\cdot)$, $\eta_R(\cdot)$, and $\eta_P(\cdot)$ map the
corresponding metrics to comparable normalized scales. Since lower BER and
lower power overhead are preferred, $1-\eta_{\mathrm{BER}}(\mathrm{BER}_t)$
and $1-\eta_P(P_{\mathrm{extra},t})$ are used as the reliability and
energy-efficiency scores. The weight vector
$\bm{w}=[w_{\mathrm{BER}},w_R,w_P]$ is selected according to the user intent.
% After executing action $\bm{a}_t$, the communication simulator evaluates
% the resulting link performance and returns several metrics, including BER,
% achievable rate, and power compensation. The scalar reward is defined as a
% weighted multi-objective utility:
% \begin{equation}
% r_t =
% w_{\text{BER}}(I_t) R_{\text{BER}}
% +
% w_{\text{rate}}(I_t) R_{\text{rate}}
% +
% w_{\text{power}}(I_t) R_{\text{power}},
% \end{equation}
% where $R_{\text{ber}}$, $R_{\text{rate}}$, and $R_{\text{power}}$
% represent normalized performance scores for reliability, throughput, and
% energy efficiency, respectively. Here, $I_t$ denotes the user intent at
% decision step $t$, and the intent-dependent reward-weight vector is
% defined as:
% \begin{equation}
% \bm{w}(I_t)
% =
% \left[
% w_{\text{ber}}(I_t),
% w_{\text{rate}}(I_t),
% w_{\text{power}}(I_t)
% \right],
% \end{equation}
% with $w_{\text{ber}}(I_t)+w_{\text{rate}}(I_t)+w_{\text{power}}(I_t)=1$.

% For the three user intents considered in this paper, the reward weights
% are set as:
% \begin{equation}
% \bm{w}(I_t)
% =
% \begin{cases}
% (0.1,\,0.8,\,0.1), 
% & I_t=\text{High-throughput},\\
% (0.8,\,0.1,\,0.1), 
% & I_t=\text{High-reliability},\\
% (0.1,\,0.1,\,0.8), 
% & I_t=\text{Energy saving}.
% \end{cases}
% \end{equation}

Since the considered problem corresponds to a single-step decision process, the policy learning objective reduces to maximizing the expected reward under the policy:
\begin{equation}
J(\theta)
=
\mathbb{E}_{\bm{s}_t\sim p_{\mathcal{S}},\, \bm{a}_t\sim\pi_\theta(\cdot|\bm{s}_t)}
\left[
r(\bm{s}_t,\bm{a}_t)
\right],
\end{equation}
where $p_{\mathcal{S}}$ denotes the state distribution induced by the generated CSI samples and user-intent prompts.

\begin{algorithm}[t]
\caption{Stage I: SFT with Heuristic Expert Samples}
\label{alg:sft}
\DontPrintSemicolon
\SetAlgoLined
\KwIn{
State distribution $p_{\mathcal{S}}$, action spaces $\{\mathcal{A}_i\}_{i=1}^{n}$, \\
replay buffer $\mathcal{D}$, 
policy network $\pi_\theta$, number of sampled states $N_{\mathrm{samp}}$, 
number of SFT epochs $N_{\mathrm{SFT}}$
}
\KwOut{Initialized policy $\pi_{\theta_{0}}$}

Initialize replay buffer $\mathcal{D}\leftarrow \emptyset$\;

\For{$t=1$ \KwTo $N_{\mathrm{samp}}$}{
    Sample $\bm{s}_t=\{\bm{H}_t,\bm{T}_t\}\sim p_{\mathcal{S}}$\;
    Randomly initialize a complete action
    $\bm{a}_t^{0}=[a_{t,1}^{0},a_{t,2}^{0},\ldots,a_{t,n}^{0}]$\;

    \For{$i=1$ \KwTo $n$}{
        \ForEach{$a_{t,i}\in\mathcal{A}_i$}{
            Evaluate
\[ \begin{aligned} r_i(a_{t,i}) &= r\!\Bigl( \bm{s}_t, \bigl[ a_{t,1}^{*},\ldots,a_{t,i-1}^{*}, a_{t,i}, \\ &\hspace{3.2em} a_{t,i+1}^{0},\ldots,a_{t,n}^{0} \bigr] \Bigr) \end{aligned} \]
        }
        Select greedy expert sub-action
        \[
        a_{t,i}^{*}
        =
        \arg\max_{a_{t,i}\in\mathcal{A}_i}
        r_i(a_{t,i})
        \]
    }

    Form expert action $
    \bm{a}_t^{*}=[a_{t,1}^{*},a_{t,2}^{*},\dots,a_{t,n}^{*}]
    $\;
    Store $(\bm{s}_t,\bm{a}_t^{*})$ in replay buffer $\mathcal{D}$\;
}

\For{$e=1$ \KwTo $N_{\mathrm{SFT}}$}{
    Sample a minibatch $\{(\bm{s},\bm{a}^{*})\}$ from $\mathcal{D}$\;
    Compute the teacher-forcing loss $\mathcal{L}_{\mathrm{SFT}}$\;
    Update $\theta$ by gradient descent on $\mathcal{L}_{\mathrm{SFT}}$\;
}

Set $\theta_{0}\leftarrow \theta$\;
\Return{$\pi_{\theta_{0}}$}\;

\end{algorithm}

% ======================================================================

\subsection{Two-Stage Training}

To efficiently train the multimodal decision model under high-dimensional discrete action spaces, we adopt a two-stage training framework consisting of supervised fine-tuning (SFT) with heuristic samples and reinforcement learning via PPO.

In the first stage, we construct a high-quality dataset using a
greedy heuristic search algorithm. The key idea is to generate an
expert-like structured action by sequentially selecting the sub-action
of each physical-layer module so as to locally maximize the communication
utility. Specifically, given a sampled state $\bm{s}_t$, we randomly initialize a
complete action
$\bm{a}_t^{0}=[a_{t,1}^{0},a_{t,2}^{0},\ldots,a_{t,n}^{0}]$.
At the $i$-th module-wise decision step, each candidate sub-action
$a_{t,i}\in\mathcal{A}_i$ of the $i$-th PHY module is evaluated by
forming a complete action sequence. Specifically, the greedy sub-action
is selected by:
\begin{equation}
a_{t,i}^{*}
=
\arg\max_{a_{t,i}\in\mathcal{A}_i}
r\!\left(
\bm{s}_t,
\left[
a_{t,1}^{*},\ldots,a_{t,i-1}^{*},
a_{t,i},
a_{t,i+1}^{0},\ldots,a_{t,n}^{0}
\right]
\right),
\end{equation}
where $a_{t,1}^{*},\ldots,a_{t,i-1}^{*}$ denote the greedy sub-actions
selected in the previous module-wise decision steps.
%After selecting $a_{t,i}^{*}$, the complete action is updated as:
%\begin{equation}
%\bm{a}_t^{i}
%=
%\left[
%a_{t,1}^{*},\ldots,a_{t,i}^{*},
%a_{t,i+1}^{0},\ldots,a_{t,n}^{0}
%\right].
%\end{equation}
By proceeding sequentially over all action dimensions, the greedy algorithm generates the expert-like action
\begin{equation}
\bm{a}_t^{*}
=
[a_{t,1}^{*},a_{t,2}^{*},\ldots,a_{t,n}^{*}].
\end{equation}
which avoids enumerating the full joint action space and significantly reduces the search complexity.

The resulting expert-like tuples $(\bm{s}_t,\bm{a}_t^{*})$ are stored in a replay buffer $\mathcal{D}$. The policy is then trained by teacher forcing to imitate the heuristic decisions. Since the policy is factorized sequentially, the SFT loss is
\begin{equation}
\mathcal{L}_{\text{SFT}}
=
-
\mathbb{E}_{(\bm{s}, \bm{a}^{*}) \sim \mathcal{D}}
\left[
\sum_{i=1}^{n}
\log \pi_{\theta}
\left(
a_i^{*} \mid \bm{s}, a_{1:i-1}^{*}
\right)
\right].
\end{equation}
This teacher-forcing objective trains each sub-action predictor conditioned on expert action prefixes, providing a strong initialization for subsequent reinforcement learning.

\begin{algorithm}[t]
\caption{Stage II: PPO Fine-Tuning with Approximate Entropy Estimate}
\label{alg:ppo}
\DontPrintSemicolon
\SetAlgoLined
\KwIn{
Initial policy $\pi_{\theta_{0}}$, state distribution $p_{\mathcal{S}}$, 
action spaces $\{\mathcal{A}_i\}_{i=1}^{n}$, simulator $\mathrm{Env}(\cdot)$, 
clipping factor $\delta$, entropy coefficient $\lambda$, 
PPO iterations $N_{\mathrm{PPO}}$
}
\KwOut{Fine-tuned policy $\pi_{\theta^*}$}

Initialize $\theta \leftarrow \theta_{0}$\;

\For{$t=1$ \KwTo $N_{\mathrm{PPO}}$}{
    Collect on-policy samples using $\pi_{\theta_{t-1}}$\;

    \ForEach{sampled state $\bm{s}_t \sim p_{\mathcal{S}}$}{
        Initialize action prefix $a_{t,1:0} \leftarrow \emptyset$\;

        \For{$i=1$ \KwTo $n$}{
            Sample sub-action 
            $a_{t,i} \sim \pi_{\theta_{t-1}}
            (a_{t,i}\mid \bm{s}_t,a_{t,1:i-1})$\;
        }

        Form $\bm{a}_t=[a_{t,1},a_{t,2},\ldots,a_{t,n}]$\;
        Execute $\bm{a}_t$ in the simulator and obtain 
        $r_t=r(\bm{s}_t,\bm{a}_t)$\;
        Set advantage $A_t \leftarrow r_t$\;
    }

    \ForEach{collected sample $(\bm{s}_t,\bm{a}_t,r_t,A_t)$}{
        Compute probability ratio $\rho_t(\theta)$\;
        Compute clipped PPO objective $\mathcal{L}_{\mathrm{PPO}}$\;
        Compute entropy estimate $\widehat{\mathcal{H}}_t$ under $\pi_{\theta}$\;
    }

    Form the total loss
    $\mathcal{L}_{\mathrm{total}}
    =
    -\mathcal{L}_{\mathrm{PPO}}
    -
    \lambda\mathbb{E}[\widehat{\mathcal{H}}_t]$\;

    Obtain $\theta_t$ by minimizing $\mathcal{L}_{\mathrm{total}}$\;
}

Set $\theta^*\leftarrow\theta$\;
\Return{$\pi_{\theta^*}$}\;

\end{algorithm}

After supervised initialization, the model is further optimized using PPO to improve performance under dynamic channel conditions. Given samples collected under the current policy $\pi_{\theta}$, PPO maximizes the clipped surrogate objective
\begin{equation}
\mathcal{L}_{\text{PPO}}
=
\mathbb{E}
\left[
\min
\left(
\rho_t(\theta) A_t,
\text{clip}(\rho_t(\theta), 1-\delta, 1+\delta) A_t
\right)
\right],
\end{equation}
where 
\begin{equation} \rho_t(\theta) = \frac{\pi_{\theta}(\bm{a}_t | \bm{s}_t)} {\pi_{\theta_{t-1}}(\bm{a}_t | \bm{s}_t)}, \end{equation} 
and $\pi_{\theta_{t-1}}$ denotes the policy from last training iteration. $\delta$ is the clipping factor that limits the deviation of the updated policy from the previous policy. Since the considered problem is a single-step decision process, the
advantage is directly given by the immediate reward, i.e.,
$A_t = r(\bm{s}_t,\bm{a}_t)$.

Since $\bm{a}_t$ is generated sequentially over a large multidimensional discrete space, computing the exact policy entropy
$\mathcal{H}(\pi_{\theta}(\cdot|\bm{s}_t))$ would require enumeration over all joint action combinations.
To encourage exploration while keeping training computationally efficient, we use a sampled conditional-entropy estimate. For the sequentially factorized policy
$\pi_{\theta}(\bm{a}_t|\bm{s}_t)=\prod_{i=1}^{n}\pi_{\theta}(a_{t,i}|\bm{s}_t,a_{t,1:i-1})$,
the entropy estimate is
\begin{equation}
\widehat{\mathcal{H}}_t
=
\sum_{i=1}^{n}
\mathcal{H}^{(i)}_{\theta}\!\left(\bm{s}_t, a_{t,1:i-1}\right),
\end{equation}
where
\begin{equation}
\begin{aligned}
\mathcal{H}^{(i)}_{\theta}\!\left(\bm{s}_t, a_{t,1:i-1}\right)
&=
-\sum_{a \in \mathcal{A}_i}
\pi_{\theta}\!\left(a \mid \bm{s}_t, a_{t,1:i-1}\right)
\\
&\quad \times
\log
\pi_{\theta}\!\left(a \mid \bm{s}_t, a_{t,1:i-1}\right).
\end{aligned}
\end{equation}
This estimator only requires the per-step categorical distribution produced by the policy network, and avoids explicit enumeration of the full joint action space.

The total loss minimized during PPO fine-tuning is
\begin{equation}
\mathcal{L}_{\text{total}}
=
-
\mathcal{L}_{\text{PPO}}
-
\lambda \,\mathbb{E}\!\left[\widehat{\mathcal{H}}_t\right],
\end{equation}
where $\lambda$ controls the strength of entropy regularization.

% This preference-aware reward formulation ensures that the optimization process is aligned with user intent, enabling adaptive trade-offs between reliability, efficiency, and computational cost.

% To further enhance the interpretability and consistency of the inferred preference vector, a preference classification loss is introduced. Specifically, the LLM-generated preference representation $\bm{p}$ is mapped to a predicted user preference label $\hat{y}{\text{pref}}$ through a lightweight classifier. The loss is defined as the cross-entropy between the predicted preference distribution and the ground-truth user preference label $y{\text{pref}}$:
% \begin{equation}
% \mathcal{L}_{\text{pref}} = - \sum{k=1}^{K} y_{\text{pref}}^{(k)} \log \hat{y}_{\text{pref}}^{(k)},
% \end{equation}
% where $K$ denotes the number of user preference categories.

% During both stages, the model leverages CoT reasoning to generate interpretable intermediate representations that reflect user preferences before producing final actions. These intermediate thoughts are supervised using the cloned samples to enhance the consistency between reasoning and decision outcomes. As training proceeds, the model evolves from mimicking expert actions to autonomous optimization, ultimately achieving robust communication decision-making under dynamic CSI and diverse user intents.

\section{Numerical Results}
In this section, we first introduce the experimental setup, then compare AGENS with representative link construction baselines, followed by ablation studies and a human--machine interaction example to evaluate its effectiveness, adaptability, and flexibility.

\begin{figure*}[t]
    \centering
    \begin{subfigure}[t]{0.32\textwidth}
        \centering
        \includegraphics[width=\linewidth]{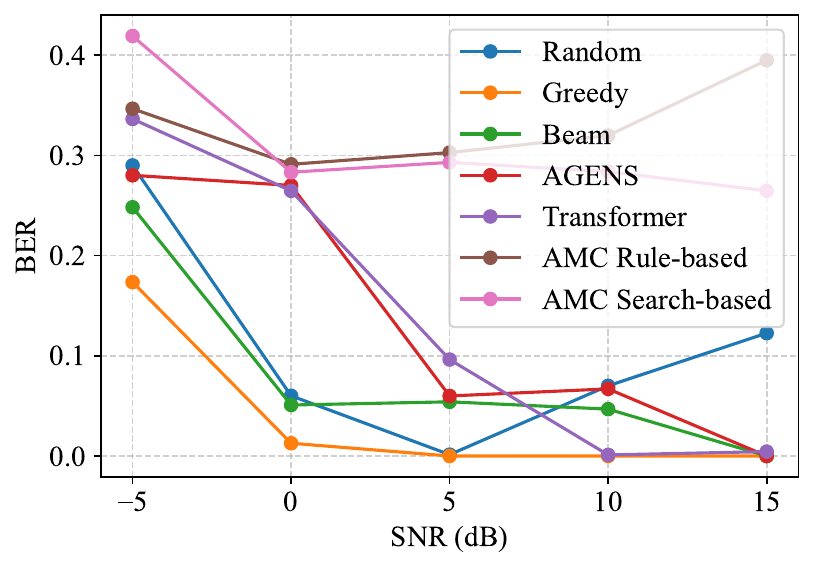}
        \caption{BER}
    \end{subfigure}
    \hfill
    \begin{subfigure}[t]{0.32\textwidth}
        \centering
        \includegraphics[width=\linewidth]{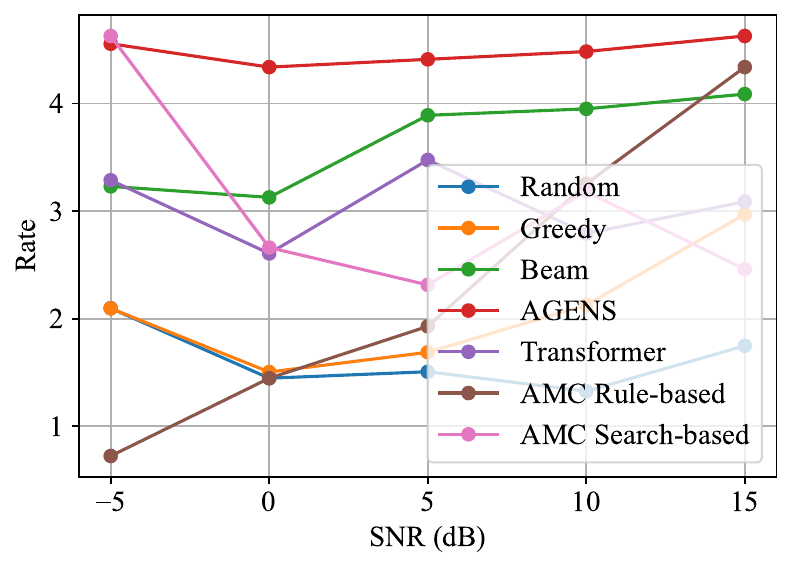}
        \caption{Rate }
    \end{subfigure}
    \hfill
    \begin{subfigure}[t]{0.32\textwidth}
        \centering
        \includegraphics[width=\linewidth]{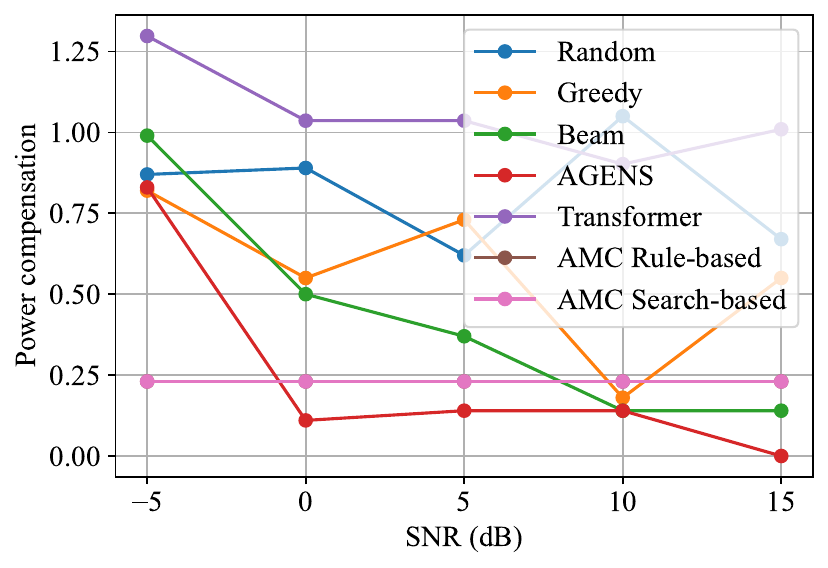}
        \caption{Power compensation}
    \end{subfigure}
    \caption{Performance comparison of different strategy selection methods under the high-throughput intent.}
    \label{fig4}
\end{figure*}

\begin{figure*}[t]
    \centering
    \begin{subfigure}[t]{0.32\textwidth}
        \centering
        \includegraphics[width=\linewidth]{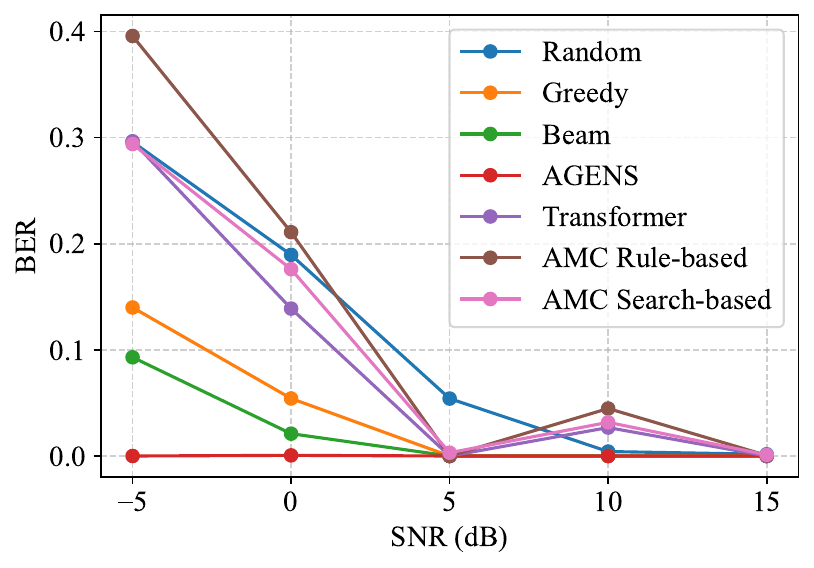}
        \caption{BER}
    \end{subfigure}
    \hfill
    \begin{subfigure}[t]{0.32\textwidth}
        \centering
        \includegraphics[width=\linewidth]{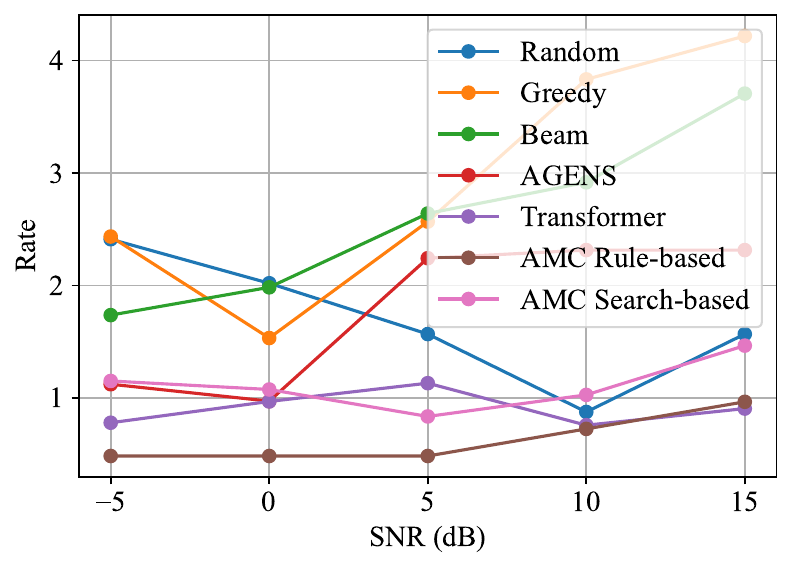}
        \caption{Rate }
    \end{subfigure}
    \hfill
    \begin{subfigure}[t]{0.32\textwidth}
        \centering
        \includegraphics[width=\linewidth]{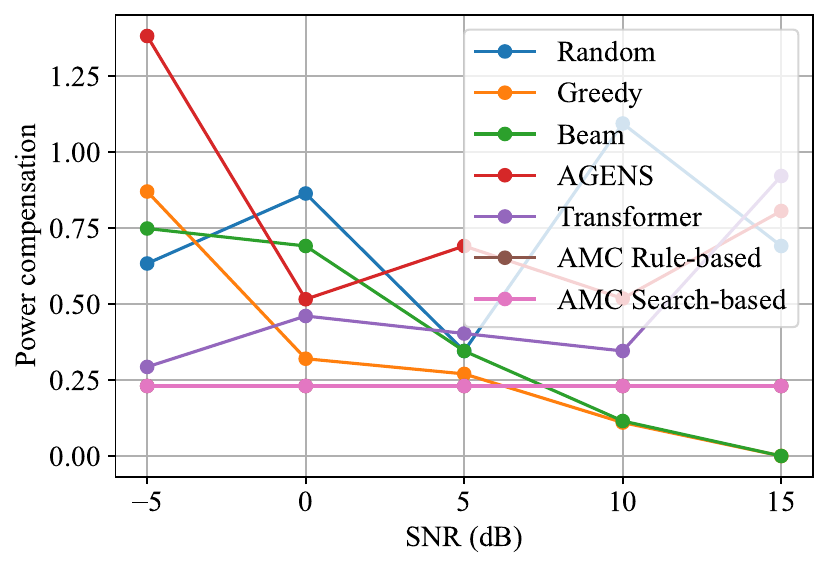}
        \caption{Power compensation}
    \end{subfigure}
    \caption{Performance comparison of different strategy selection methods under the high-reliability intent.}
    \label{fig5}
\end{figure*}

\begin{figure*}[t]
    \centering
    \begin{subfigure}[t]{0.32\textwidth}
        \centering
        \includegraphics[width=\linewidth]{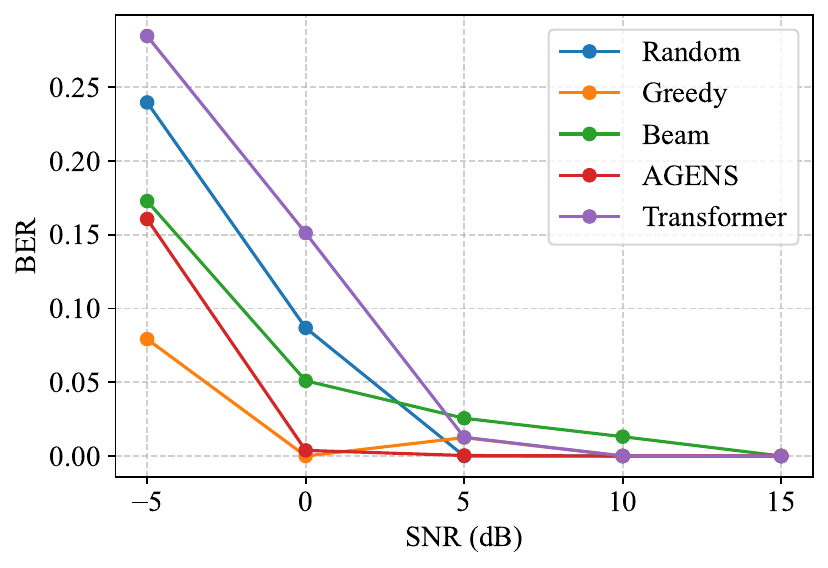}
        \caption{BER}
    \end{subfigure}
    \hfill
    \begin{subfigure}[t]{0.32\textwidth}
        \centering
        \includegraphics[width=\linewidth]{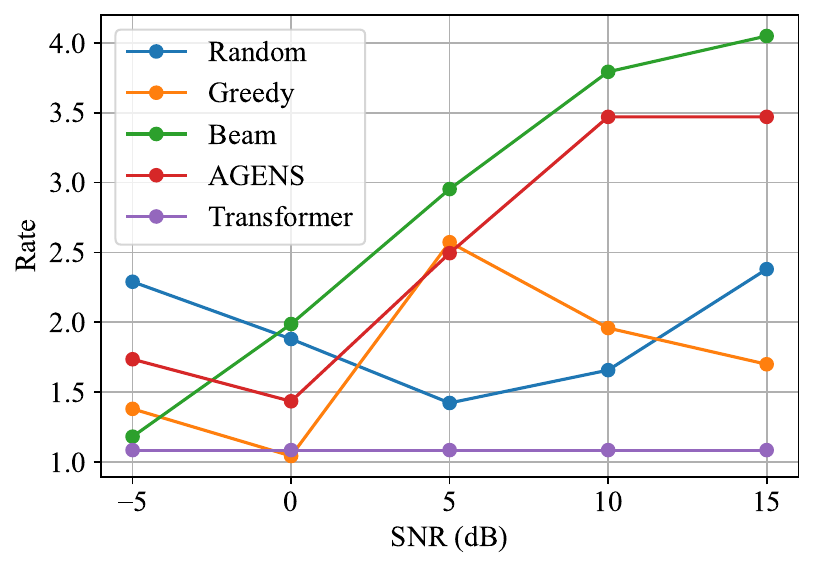}
        \caption{Rate }
    \end{subfigure}
    \hfill
    \begin{subfigure}[t]{0.32\textwidth}
        \centering
        \includegraphics[width=\linewidth]{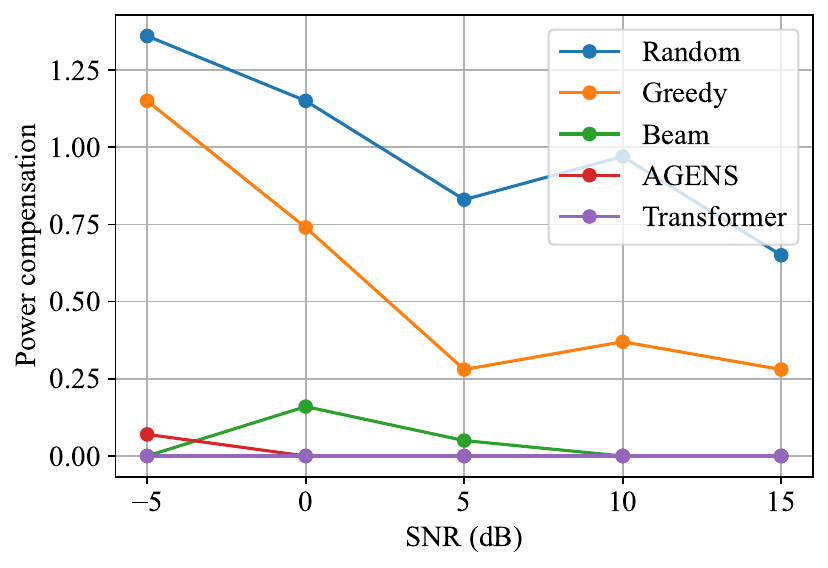}
        \caption{Power compensation}
    \end{subfigure}
    \caption{Performance comparison of different strategy selection methods under the energy-saving intent.}
    \label{fig6}
\end{figure*}
\subsection{Experimental Setup}\label{AA}
\begin{table}[t]
\centering
\caption{Key simulation parameters for CSI generation}
\label{tab:csi_generation_params}
\renewcommand{\arraystretch}{1.2}
\begin{tabular}{c|c}
\hline
\textbf{Parameter} & \textbf{Value} \\
\hline

Number of transmitters & $1$ \\
Number of receivers & $1$ \\
Number of BS antennas & $4$ \\
Number of UE antennas & $2$ \\
Number of streams per transmitter & $2$ \\
Subcarrier spacing & $30$ kHz \\
FFT size & $128$ \\
Cyclic prefix length & $10$ \\
Number of OFDM symbols & $14$ \\
Pilot OFDM symbol indices & $\{2,9\}$ \\
CIR batch size & $10$ \\
Target CIRs per configuration & $50$ \\
\hline
\end{tabular}
\end{table}
We generate CSI datasets with diverse parameter configurations using Sionna-RT, covering heterogeneous channel conditions and geometric propagation scenarios. The key simulation parameters for CSI generation are summarized in Table~\ref{tab:csi_generation_params}. For each CSI sample, a natural-language user-intent description is assigned so that channel conditions and user requirements vary jointly. The dataset is constructed under three urban city scenarios, namely Munich, Etoile, and Florence, and is split into training, validation, and test sets containing 12,000, 2,000, and 4,000 samples, respectively.

We consider three representative 6G service preferences: high throughput, high reliability, and energy-efficient. In the reward function defined in Section~V.A, the intent-dependent weight vector $\bm{w}=[w_{\mathrm{BER}},w_R,w_P]$ is set as: \begin{equation} \bm{w} = \begin{cases} [0.1,\,0.8,\,0.1], & \text{high-throughput},\\ [0.8,\,0.1,\,0.1], & \text{high-reliability},\\ [0.1,\,0.1,\,0.8], & \text{energy-saving}. \end{cases} \end{equation} For model implementation, GPT-2 Medium with 355M parameters is adopted as the LLM backbone to construct the intelligent agent. The candidate configuration options of different PHY-layer modules are summarized in Table~\ref{tab:phy_config_options}. 
\begin{table}[t]
\centering
\caption{Candidate configuration options for PHY-layer modules.}
\label{tab:phy_config_options}
\begin{tabular}{l l}
\hline
\textbf{PHY-layer module} & \textbf{Candidate options} \\
\hline
Modulation & QPSK, 16QAM, 64QAM, 256QAM \\

Coding rate & $1/3$, $1/2$, $2/3$, $3/4$, $4/5$ \\

Power compensation(dB) & 0, 1, 2, 3, 4, 6, 8  \\
Channel coding & LDPC, Polar, Convolutional code \\
Channel estimation & LS, MMSE \\
Equalization & LMMSE, MFE, ZF \\
Precoding & ZF, RZF \\
\hline
\end{tabular}
\end{table}
The pretrained language model is mainly employed for language understanding and user-intent inference. Therefore, he proposed framework is not tied to a specific LLM model and can be replaced with other LLMs. 
% Conventional AMC is not adopted as a direct baseline because it mainly selects modulation and coding schemes according to channel quality, whereas our task requires intent-aware full-link construction across multiple coupled PHY-layer modules.  Such a high-dimensional multi-objective decision problem cannot be directly covered by standard AMC rules or lookup tables. In addition, AMC thresholds are typically calibrated for a fixed system setting and a specific optimization objective. They are difficult to generalize when the service preference, channel scenario, or candidate configuration space changes. Therefore, we compare AGENS with random selection, greedy search, beam search, and a Transformer-based variant, which share the same structured action space and are more suitable baselines for evaluating full-link strategy generation.
\subsection{Model Decision Performance}\label{BB}
We first evaluate the capability of AGENS to generate personalized communication strategies under different user preferences. Four baseline methods are considered for comparison: 
(i) \emph{Random}, which independently samples each PHY-module configuration from its candidate set; 
(ii) \emph{Greedy Search}, which sequentially selects the sub-action with the highest immediate reward conditioned on previously selected sub-actions; 
(iii) \emph{Beam Search}, which maintains the top-2 partial strategies at each decision step and expands them until a complete PHY-layer configuration is obtained; and 
(iv) \emph{Transformer-based AGENS}, where the LLM backbone is replaced by a standard Transformer encoder while the rest of the decision-making framework remains unchanged. 
For the random baseline, different random seeds are used under different user preferences to ensure sufficient diversity in the generated strategies.
To further compare AGENS with conventional link adaptation schemes, we additionally introduce two AMC-based baselines:
(v) \emph{AMC Rule-based}, which selects the modulation and coding scheme according to an SNR-to-MCS rule derived from extensive link-level experiments. Specifically, for each SNR region, the rule selects the highest-rate MCS among the candidates that satisfy the target BER requirement; 
(vi) \emph{AMC Search-based}, which searches among candidate modulation and coding schemes according to the corresponding intent-oriented objective. AMC Search-based is used as an oracle-style upper bound within the AMC decision space rather than a practical online AMC algorithm. Since AMC only adapts modulation and coding, it mainly affects throughput and reliability but cannot directly control power compensation. Thus, AMC-based baselines are evaluated only under the high-throughput and high-reliability intents.

Under the high-throughput intent, AGENS adaptively favors higher-rate configurations. As shown in Fig.~\ref{fig4}, AGENS achieves the highest $R$ across most SNR regimes, especially at medium and high SNR values, while keeping BER within an acceptable range. Greedy search and beam search can obtain relatively low BER in some cases, but their achievable rates are limited by local prefix-wise optimization and therefore do not fully exploit the full-link action space. The AMC-based baselines also exhibit clear limitations. AMC Rule-based follows fixed SNR-to-MCS thresholds; although its rate increases as SNR improves, the resulting BER can remain high when the selected MCS is overly aggressive. AMC Search-based serves as an oracle within the AMC decision space, but it optimizes only modulation and coding while keeping the other PHY modules fixed. As a result, AMC-based methods cannot consistently preserve a favorable throughput--reliability trade-off across different SNR values. In contrast, AGENS jointly coordinates all configurable PHY-layer modules and optimizes the end-to-end intent-oriented reward, thereby better satisfying the high-throughput requirement.

Under the high-reliability intent, AGENS effectively captures the requirement for BER minimization and generates more robust communication strategies. As shown in Fig.~\ref{fig5}, AGENS achieves nearly zero BER across the entire SNR range by selecting more conservative PHY-layer configurations. This reliability gain comes at the cost of a lower transmission rate, reflecting an explicit trade-off between spectral efficiency and link robustness. Greedy search and beam search can also reduce BER at medium-to-high SNR values, but their performance degrades under poor channel conditions due to their reliance on local optimization. Similarly, AMC Rule-based and AMC Search-based schemes improve reliability when SNR is sufficiently high, yet their BER remains high in low-SNR regimes. 

%When minimizing BER is the primary objective, AGENS effectively captures the reliability-oriented intent and generates more robust communication strategies. As shown in Fig.~\ref{fig5}, AGENS achieves nearly zero BER across the whole SNR range by selecting more conservative PHY-layer configurations. This reliability improvement is accompanied by a reduced transmission rate, which reflects an intentional trade-off between spectral efficiency and link robustness. Greedy and beam search can also reduce BER at medium-to-high SNR values, but they still suffer from noticeable degradation under poor channel conditions. Similarly, both AMC Rule-based and AMC Search-based schemes improve reliability when the SNR is sufficiently high, but their BER remains high at low SNR values. This further indicates that optimizing only modulation and coding is insufficient to ensure robust end-to-end performance under challenging channel conditions. By comparison, AGENS benefits from reinforcement-learning-based feedback and LLM-based reasoning, enabling coordinated decisions over multiple PHY modules rather than relying solely on modulation and coding adaptation.

Under the energy-saving intent, AGENS tends to select configurations with lower power compensation, thereby reducing the additional transmit-power overhead. To maintain link reliability under constrained power usage, it adopts more conservative modulation and coding choices, resulting in a balanced trade-off among power consumption, BER, and achievable rate. In contrast, the random baseline lacks structured decision logic, while greedy search and beam search are limited by local module-level optimization and may fail to jointly satisfy power-efficiency and end-to-end performance requirements.
\begin{figure*}[t]
    \centering
    \begin{subfigure}[t]{0.32\textwidth}
        \centering
        \includegraphics[width=\linewidth]{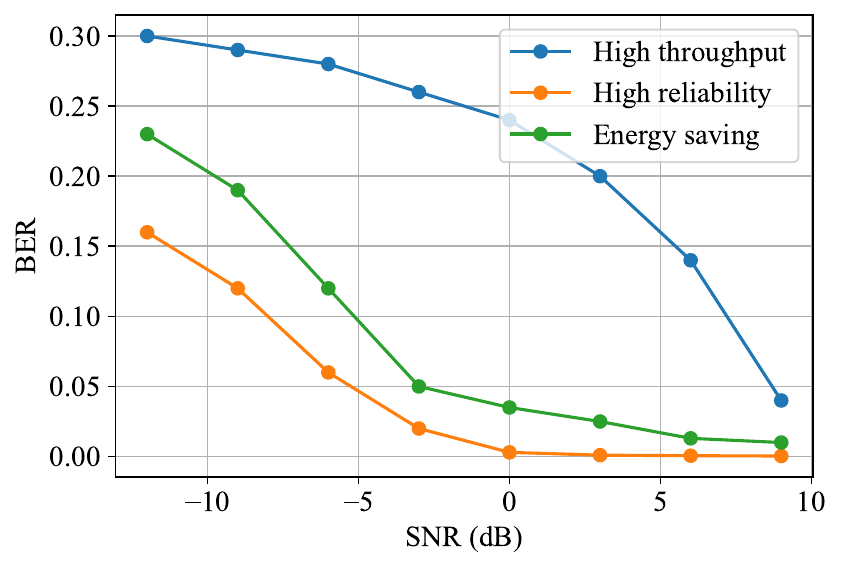}
        \caption{BER}
    \end{subfigure}
    \hfill
    \begin{subfigure}[t]{0.32\textwidth}
        \centering
        \includegraphics[width=\linewidth]{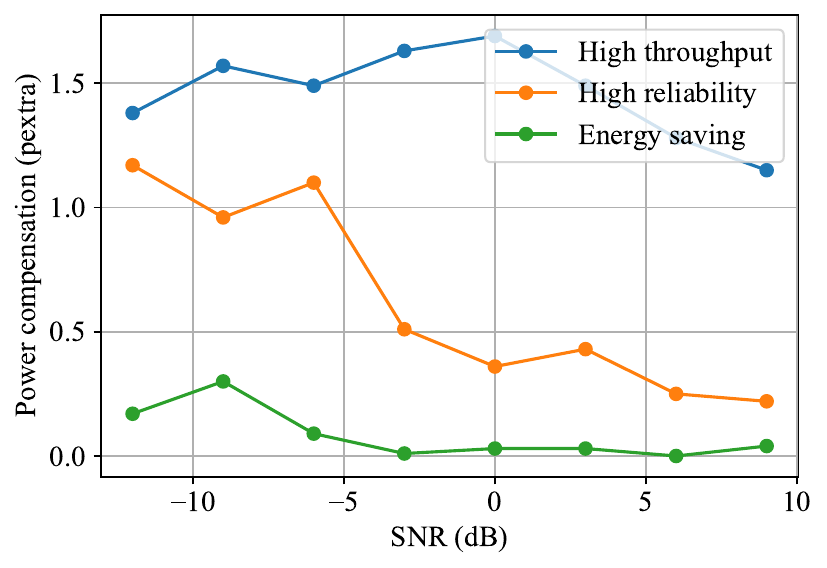}
        \caption{Power compensation }
    \end{subfigure}
    \hfill
    \begin{subfigure}[t]{0.32\textwidth}
        \centering
        \includegraphics[width=\linewidth]{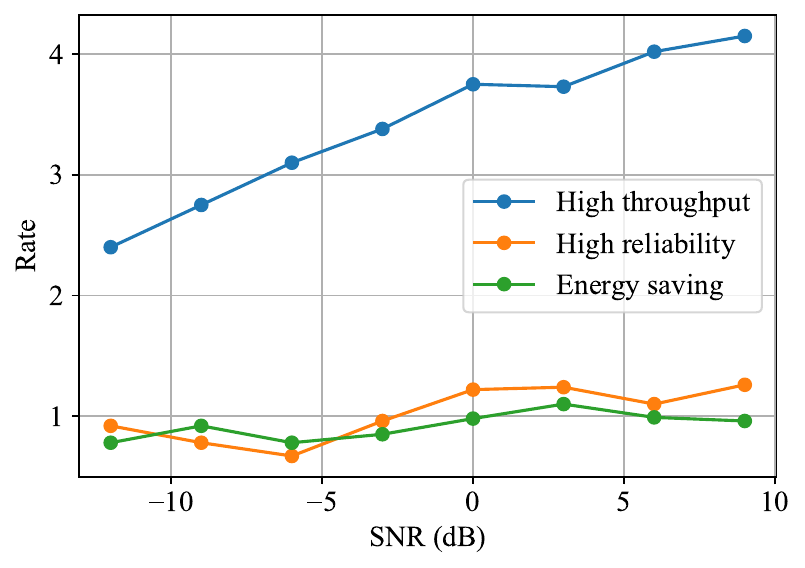}
        \caption{Rate}
    \end{subfigure}
    \caption{Performance comparison under three types of user intents.}
    \label{fig7}
\end{figure*}

%It can also be observed that the Transformer-based method performs worse than traditional heuristic approaches when either high throughput or high reliability is prioritized. Although it achieves relatively low power consumption under the energy efficiency objective, it often produces overly aggressive strategies. As shown in Fig.~\ref{fig6}, the Transformer-based method tends to generate rigid decisions, selecting nearly identical transmission rates across all scenarios. Such strategies result in excessively high BER, which may render the communication system unreliable in practice. These observations highlight the necessity of incorporating LLMs. 
The Transformer-based variant performs worse than the heuristic approaches when either high throughput or high reliability is prioritized. Although it achieves relatively low power consumption under the energy-saving intent, it often produces overly aggressive and less adaptive strategies. As shown in Fig.~\ref{fig6}, the Transformer-based variant tends to make rigid decisions, selecting nearly identical transmission rates across different scenarios. These strategies lead to excessively high BER and may therefore make the communication link unreliable in practice. This result highlights the importance of incorporating an LLM backbone for intent-aware and channel-adaptive strategy generation.

It is worth noting that some performance curves, by random or Greedy baseline,  exhibit noticeable fluctuations. This is partly due to the inherent randomness of these strategies and partly because they rely on local optimization of individual modules, which may lead to instability in the overall performance of multi-module communication systems. In contrast, AGENS produces smoother curves across different metrics, indicating better global coordination capability and higher decision stability.

Furthermore, we evaluate AGENS under different user intents with identical channel conditions. As shown in Fig.~\ref{fig7}, AGENS produces distinct performance trends for the three intents, confirming its intent-awareness. The reliability-oriented strategy minimizes BER with more conservative configurations, the throughput-oriented strategy achieves a higher rate by accepting increased BER and power usage, and the energy-aware strategy reduces power compensation while maintaining moderate rate and reliability.
% \begin{figure*}[t]
%     \centering
%     % 第一张图
%     \begin{minipage}[t]{0.32\textwidth}
%         \centering
%         \includegraphics[width=\textwidth]{BER_normal.pdf}
%         \caption*{(a) BER }
%     \end{minipage}
%     \hfill
%     % 第二张图
%     \begin{minipage}[t]{0.32\textwidth}
%         \centering
%         \includegraphics[width=\textwidth]{Rate_normal.pdf}
%         \caption*{(b)  Rate }
%     \end{minipage}
%     \hfill
%     % 第三张图
%     \begin{minipage}[t]{0.32\textwidth}
%         \centering
%         \includegraphics[width=\textwidth]{Complexity_normal.pdf}
%         \caption*{(c) Complexity  }
%     \end{minipage}
%     \caption{Performance comparison of different strategy selection methods for general strategy under various SNRs.}
%     \label{fig6}
% \end{figure*}

\begin{figure}[htbp]
    \centering
    \includegraphics[width=0.47\textwidth]{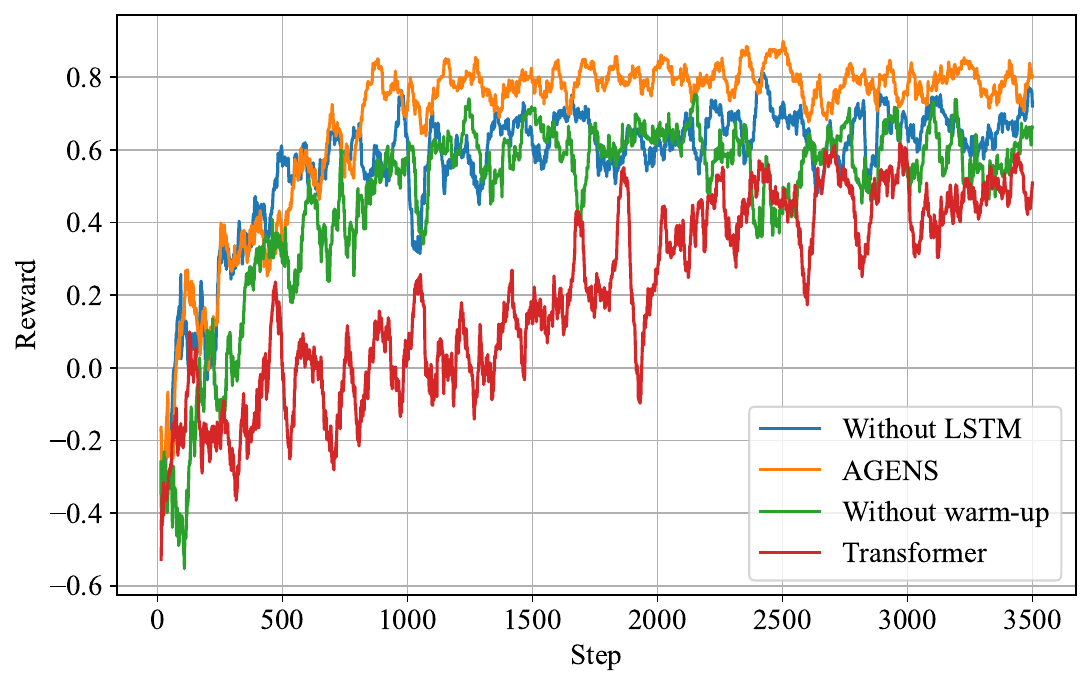} 
    \caption{Training reward curves for different model variants}
    \label{fig8}
\end{figure}

\subsection{Adaptation to Heterogeneous Channel Scenarios}
\label{subsec:channel_awareness}

This subsection provides a qualitative analysis of whether the proposed AGENS framework can adapt its PHY-layer strategy to heterogeneous channel environments while satisfying the \emph{throughput-oriented} user intent.
The user preference label is not directly provided to the policy network. Instead, each preference is expressed as a natural-language prompt and encoded by the language model. Therefore, AGENS infers the user intent from the text prompt and jointly considers CSI and SNR to generate a discrete multi-module strategy.

We consider two representative channel scenarios with distinct propagation characteristics:
\begin{itemize}
    \item \textbf{Scenario-A (simple LoS-like)}: near distance (\SI{10}{m}--\SI{60}{m}), medium path gain (\SI{-90}{dB} to \SI{-50}{dB}), few multipath components (maximum depth $=1$), and static terminals.
    \item \textbf{Scenario-B (complex NLoS-rich)}: far distance (\SI{200}{m}--\SI{450}{m}), weak path gain only (\SI{-140}{dB} to \SI{-90}{dB}), rich multipath (maximum depth $=8$), and vehicular mobility (velocity $(60,60,20)km/h$).
\end{itemize}
These two settings represent a relatively benign near-static channel and a challenging long-range mobile rich-scattering channel, respectively.

We sample CSI inputs from both scenarios and generates corresponding user-intent prompts requesting throughput prioritization. We also use different SNRs for these two scenarios. Specifically, Scenario-A uses $\text{SNR}=5$ dB, while Scenario-B uses $\text{SNR}=-5$ dB. The generated strategies by AGENs for these two senarios are:
\begin{align}
\mathbf{a}_A &= [3,\,2,\,5,\,0,\,1,\,0,\,0], \\
\mathbf{a}_B &= [2,\,4,\,6,\,0,\,0,\,0,\,1],
\end{align}
respectively. Table~\ref{tab:agens_strategy_ab_throughput} summarizes the per-module decisions.
Compared with Scenario-A, Scenario-B corresponds to a harsher propagation condition with a lower SNR. AGENS accordingly adjusts several key PHY modules, including modulation order, coding-rate selection, channel estimation, and precoding, while keeping other components unchanged.
This comparison demonstrates that AGENS can perform channel-aware adaptive strategy generation: it preserves the throughput-oriented intent while modifying PHY-layer configurations to accommodate heterogeneous propagation conditions.

\begin{table*}[t]
\centering
\caption{Strategies generated by AGENS for throughput-oriented intent under two channel scenarios.}
\label{tab:agens_strategy_ab_throughput}
\begin{tabular}{lcc}
\hline
\textbf{Decision module} & \textbf {Scenario-A} ($\text{SNR}=5$ dB) & \textbf{Scenario-B} ($\text{SNR}=-5$ dB) \\
\hline
Modulation ($a_{\text{mod}}$) & 256QAM & 64QAM \\
Rate control ($a_{\text{rate}}$) & 2/3 & 4/5 \\
Power control ($a_{\text{pc}}$) & 6  dB & 8 dB \\
Channel coding ($a_{\text{enc}}$) & LDPC & LDPC \\
Channel estimation ($a_{\text{ce}}$) & MMSE & LS \\
Equalization ($a_{\text{eq}}$) &  LMMSE & LMMSE \\
Precoding ($a_{\text{prec}}$) &    ZF & RZF \\
\hline
\end{tabular}
\end{table*}

\subsection{Generalization to Unseen-City Scenarios}

To further evaluate the generalization capability of AGENS, we conduct an unseen-city test on the San Francisco scenario. 
The model is directly evaluated under \(\mathrm{SNR}=2\) dB, with one representative sample selected for each of the three user intents: high-throughput, high-reliability, and energy-saving. 
For each case, the strategy generated by AGENS is compared with the strategy obtained by exhaustive search in the target environment, which serves as the reference strategy with the maximum reward. 

The results are summarized in Table~\ref{tab:unseen_city_generalization}. It can be observed that, under different user requirements, the rewards achieved by AGENS are close to those obtained by exhaustive search, indicating that AGENS can generalize to unseen city environments and produce competitive full-link configurations. 
The high-throughput case further reveals a practically meaningful behavior. Although exhaustive search achieves the highest $R$ and the maximum reward, it also leads to a relatively high BER, which may degrade communication reliability. In contrast, AGENS selects a configuration with a slightly lower $R$ but reduces BER to nearly zero by using a larger power compensation level. This result suggests that AGENS does not merely pursue the numerically optimal reward, but learns to balance throughput, reliability, and power overhead in a user-aware manner. Such behavior is desirable in practical communication systems, where maximizing a predefined metric may not always yield the best user experience. By jointly interpreting user intent and channel conditions, AGENS provides a flexible solution for adaptive full-link PHY-layer configuration in unseen environments.

\begin{table*}[t]
\centering
\caption{Results on the unseen San Francisco scenario under \(\mathrm{SNR}=2\) dB.}
\label{tab:unseen_city_generalization}
\fontsize{6pt}{6.6pt}\selectfont
\renewcommand{\arraystretch}{1.05}
\resizebox{0.7\textwidth}{!}
{
\begin{tabular}{c|c|c|c|c|c|c}
\hline
\textbf{Intent} 
& \textbf{Method}
& \textbf{Strategy}
& \textbf{BER}
& \textbf{Rate}
& \textbf{Power}
& \textbf{Reward} \\
\hline

High throughput
& Optimal 
& $[3,4,1,0,1,2,1]$
& 0.298
& 4.631
& 0.231
& 0.807 \\
\cline{2-7}
& AGENS
& $[2,4,6,0,1,0,1]$
& 0
& 3.470
& 1.842
& 0.789 \\
\hline

High-reliability
& Optimal
& $[2,4,3,0,1,2,1]$
& 0
& 3.471
& 0.691
& 0.815 \\
\cline{2-7}
& AGENS
& $[0,4,6,0,1,0,1]$
& 0
& 1.157
& 1.842
& 0.728 \\
\hline

Energy saving
& Optimal 
& $[1,1,0,0,1,2,0]$
& 0
& 1.446
& 0
& 0.904 \\
\cline{2-7}
& AGENS
& $[0,4,0,0,1,0,1]$
& 0
& 1.157
& 0
& 0.898 \\
\hline

\end{tabular}
}
\end{table*}

\subsection{Ablation Experiment}\label{CC}
In this subsection, we conduct ablation studies to investigate the effects of architectural components and training strategies on learning performance. Specifically, we compare the training reward curves under four configurations: the full AGENS model, a Transformer-based variant, a variant without the LSTM-based state modeling module, and an AGENS model trained solely with PPO without heuristic initialization.

As shown in Fig.~\ref{fig8} , removing the LSTM module leads to significantly slower convergence and lower final rewards compared to the full model. This performance degradation indicates that the LSTM-based sequential modeling effectively captures dependencies among sub-actions, enabling coordinated decision-making across PHY-layer modules rather than treating each module independently. As a result, the full model achieves stronger policy expressiveness and better global optimization performance.
When trained using PPO alone, the model can eventually reach a reward level comparable to that of the full two-stage framework; however, the convergence is substantially slower. This can be attributed to inefficient and unguided exploration during the early training stage. In contrast, heuristic-guided supervised fine-tuning provides high-quality prior knowledge, accelerates subsequent reinforcement learning, and stabilizes policy improvement. Moreover,  replacing the LLM backbone with a standard Transformer leads to both slower convergence and lower final rewards, demonstrating the importance of pretrained LLM representations for intent-aware communication strategy planning.
\subsection{Model Complexity}\label{DD}
\begin{table}[ht]
\centering
\caption{Inference time for one strategy-planning task.}
\label{TableI}
\begin{tabular}{|c|c|c|c|}
\hline

 \textbf{method} & \textbf{inference time} \\
\hline
  Random  & 0.005s \\
\hline
   AGENS & 0.167 s \\
\hline
  Beam  & 446.472s \\
\hline
  Greedy  & 307.987s \\
\hline

\end{tabular}
\end{table}
To evaluate the practical feasibility of the proposed framework, we compare the inference time of different strategy-planning methods for a single planning task. As shown in Table~\ref{TableI}, search-based heuristic methods incur substantially higher latency than AGENS because they explicitly explore or optimize over candidate actions. In contrast, AGENS generates a complete strategy through a single forward pass, significantly reducing inference time. We note that AGENS is not intended for slot-level real-time reconfiguration. Instead, it is designed for link-configuration planning at a slower control timescale, where a communication strategy is generated for a given channel condition and user intent and then maintained over an operation interval. Under this setting, AGENS demonstrates strong practical feasibility.

% In real-time physical layer link construction—where channel conditions and user requirements may change rapidly—such low-latency decision-making is critical. 
% In the ablation study, we compare the complete CoT-RL training framework with a variant that excludes the behavior cloning stage and relies solely on reinforcement learning. As illustrated in Fig.$\ref{fig7}$, when behavior cloning supervision is removed, the model struggles to maintain a proper balance between BER and throughput under the high-throughput preference setting. Consequently, the resulting strategies become overly conservative, leading to noticeably lower system throughput. These observations indicate that behavior cloning provides a crucial initialization that enables the model to learn effective preference-aware strategies. In contrast, pure reinforcement learning tends to converge to conservative policies that generalize more easily but fail to achieve optimal performance.
% \begin{figure*}[t]
%     \centering
%     % 第一张图
%     \begin{minipage}[t]{0.48\textwidth}
%         \centering
%         \includegraphics[width=\textwidth]{BER_ab.pdf}
%         \caption*{(a) BER }
%     \end{minipage}
%     \hfill
%     % 第二张图
%     \begin{minipage}[t]{0.48\textwidth}
%         \centering
%         \includegraphics[width=\textwidth]{Rate_ab.pdf}
%         \caption*{(b)  Rate }
%     \end{minipage}
%     \hfill
%     \caption{Ablation experiments.}
%     \label{fig7}
% \end{figure*}
\subsection{Interactive Intent-Aware Strategy Generation}\label{EE}
\begin{figure}[htbp]
    \centering
    \includegraphics[width=0.48\textwidth]{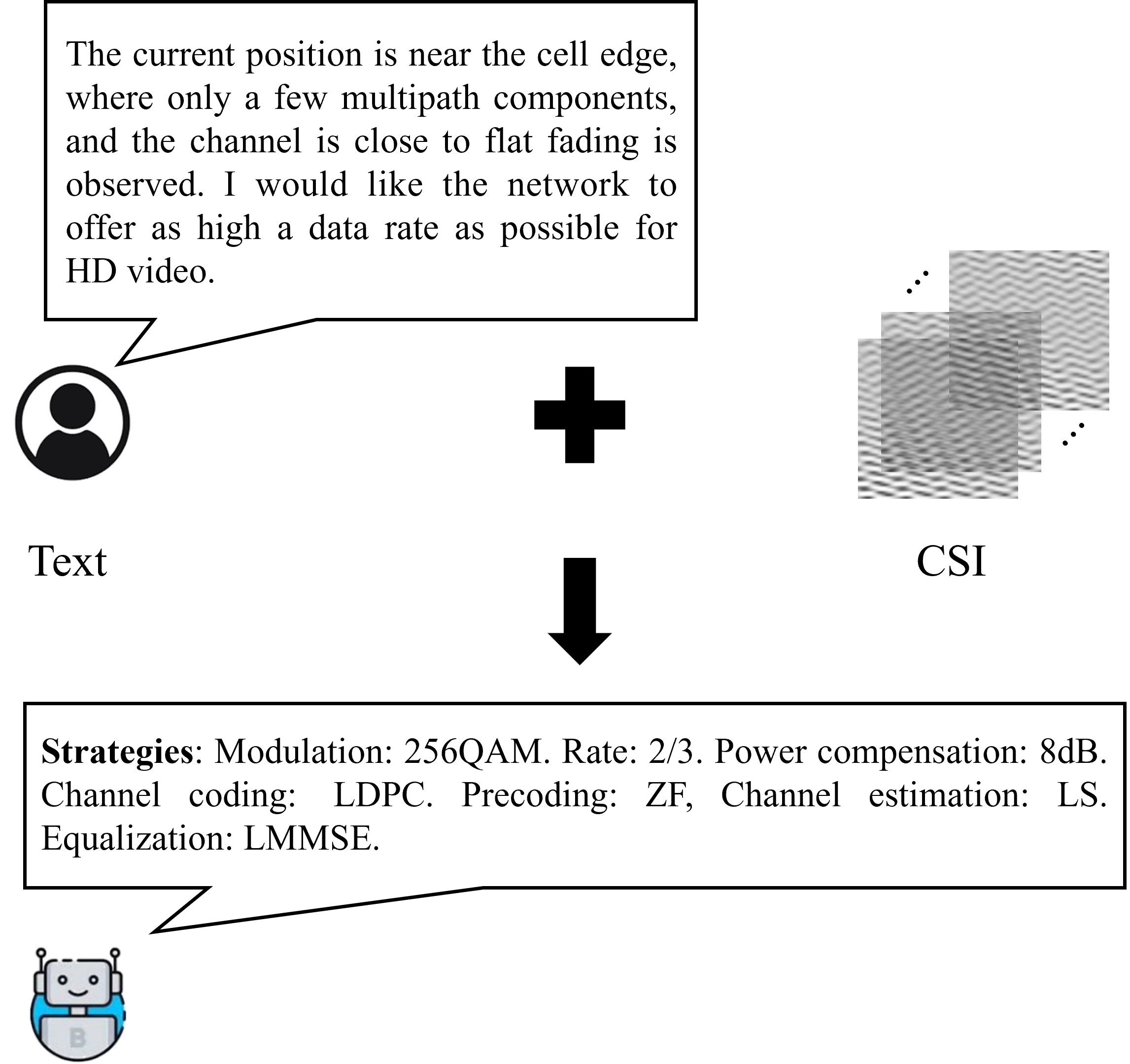} 
    \caption{Example of intent-aware interaction with AGENS.}
    \label{fig9}
\end{figure}
This subsection presents a human--machine interaction example to illustrate the practical functionality of AGENS. As shown in Fig.~\ref{fig9}, the user provides environmental information and transmission requirements as multimodal inputs. AGENS interprets the high-throughput intent, captures the underlying communication context, and generates an optimized PHY-layer configuration with interpretable decision outputs. The generated strategy is consistent with the current channel conditions and aligns with the user's service preference, such as video streaming or cloud gaming. This example demonstrates the ability of AGENS to perform semantic reasoning and adaptive strategy generation in multimodal communication scenarios.
\section{Conclusion}
This paper proposed AGENS, an agentic decision-making framework for environment- and intent-aware full-link construction in AI-native 6G communications. By jointly perceiving CSI and natural-language user intent, AGENS generates customized PHY-layer configurations across multiple communication modules. Specifically, textual instructions and CSI are first fused through a cross-modal representation module, followed by semantic and environment-aware feature modeling using a pretrained LLM backbone. An LSTM-based sequential action generator is then introduced to capture dependencies among PHY-layer sub-actions and produce structured full-link strategies. For training, we developed a two-stage framework in which expert-like samples generated by a greedy heuristic algorithm provide stable supervised initialization, and PPO with an approximate entropy estimate further refines the policy under intent-dependent rewards.

The proposed AGENS offers \textbf{three key advantages}. First, it remains \textbf{compatible} with existing modular communication systems by selecting strategies from predefined PHY-layer module candidates, rather than requiring a fully redesigned transceiver architecture. Second, to the best of our knowledge, AGENS is the first framework that \textbf{automatically} constructs full-link PHY configurations according to both wireless environment and user intent, moving beyond conventional single-module or objective-fixed optimization. Third, AGENS demonstrates strong \textbf{adaptability} across heterogeneous channel conditions, diverse user preferences, and unseen city scenarios, generating competitive strategies without time-consuming online search. Simulation results show that AGENS achieves much lower inference latency than search-based heuristic methods while maintaining competitive link-level performance under diverse communication scenarios.

\end{document}